\documentclass[11pt]{article}
\usepackage{amsmath,amsfonts,amssymb,graphicx,epsfig,pdflscape}
\usepackage[usenames]{color}
\usepackage{pstricks}
\numberwithin{equation}{section}

\newcommand{\nc}{\newcommand}
\nc\disp{\displaystyle}
\nc{\fh}{\hat{f}}
\nc{\muh}{\hat{\mu}}
\nc{\nuh}{\hat{\nu}}
\nc{\spos}[2]{\makebox(0,0)[#1]{$\sm{#2}$}}
\nc{\sm}[1]{{\scriptstyle #1}}
\nc{\bib}{\bibitem}
\nc{\al}{\alpha}
\nc{\g}{\gamma}
\nc{\G}{\Gamma}
\nc{\D}{\Delta}
\nc{\eps}{\epsilon}
\nc{\la}{\lambda}
\nc{\La}{\Lambda}
\nc{\var}{\varphi}
\nc{\pa}{\partial}
\nc{\nn}{\nonumber \\ }
\nc{\hf}{\frac{1}{2}}
\nc{\dz}{\frac{dz}{2\pi i}}
\nc{\bin}[2]{\left(\!\!\!\begin{array}{c} {#1}\\ {#2} \end{array}\!\!\!\right)}
\nc{\be}{\begin{equation}}
\nc{\ee}{\end{equation}}
\nc{\bea}{\begin{eqnarray}}
\nc{\eea}{\end{eqnarray}}
\nc{\bra}[1]{\langle {#1}|}
\nc{\ket}[1]{|{#1}\rangle}
\nc{\ketw}[1]{({#1})^{\phantom{a}}_{{\cal W}}}
\nc{\chit}{\raisebox{0.25ex}{$\chi$}}
\nc{\chih}{\raisebox{0.25ex}{$\hat\chi$}}
\nc{\Db}{\mbox{\boldmath $D$}}
\nc{\Hb}{\mbox{\boldmath $H$}}
\nc{\calH}{{\cal H}}
\nc{\calR}{{\cal R}}
\nc{\calL}{{\cal L}}
\nc{\calV}{{\cal V}}
\nc{\Hc}{{\cal H}}
\nc{\Rc}{{\cal R}}
\nc{\Lc}{{\cal L}}
\nc{\Vc}{{\cal V}}
\nc{\Ib}{\mbox{\boldmath $I$}}
\nc{\qb}{\bar{q}}
\nc{\Ac}{\mathcal{A}}
\nc{\Bc}{\mathcal{B}}
\nc{\Cc}{\mathcal{C}}
\nc{\Dc}{\mathcal{D}}
\nc{\Ec}{\mathcal{E}}
\nc{\Ic}{\mathcal{I}}
\nc{\Oc}{\mathcal{O}}
\nc{\Xc}{\mathcal{X}}
\nc{\Yc}{\mathcal{Y}}
\nc{\Zc}{\mathcal{Z}}
\nc{\fus}{\mbox{}\,\hat\otimes\,\mbox{}}
\def\vvdots{\mathinner{\mkern1mu\raise1pt\vbox{\kern7pt\hbox{.}}\mkern2mu
  \raise4pt\hbox{.}\mkern2mu\raise7pt\hbox{.}\mkern1mu}}
\nc{\gauss}[2]{\left[\!\!\begin{array}{c} {#1}\\ {#2} \end{array}\!\!\right]}
\nc{\sbin}[2]{\left\{\!\!\!\begin{array}{c} {#1}\\ {#2} 
\end{array}\!\!\!\right\}}
\nc{\sbinlr}[2]{\Big\langle\!\!\begin{array}{c} {#1}\\ {#2} 
\end{array}\!\!\Big\rangle}
\nc{\bino}[2]{\left(\!\!\begin{array}{c} {#1}\\ {#2} \end{array}\!\!\right)}
\definecolor{lightblue}{rgb}{.61,.61,1}
\definecolor{midblue}{rgb}{.7,.7,1}
\definecolor{lightlightblue}{rgb}{.85,.85,1}
\definecolor{lightestblue}{rgb}{.96,.96,1}
\definecolor{lightpurple}{rgb}{1,.65,1}
\nc{\ch}{{\rm ch}}
\nc{\R}{{\cal R}}
\nc{\dkk}{\delta_{j,\{k,k'\}}^{(2)}}
\nc{\drr}{\delta_{j,\{r,r'\}}^{(2)}}
\nc{\ddkk}{\delta_{j,\{k,k'\}}^{(4)}}
\nc{\dddkk}{\delta_{j,\{k,k'\}}^{(8)}}
\nc{\dnn}{\delta_{j,\{n,n'\}}^{(2)}}
\nc{\ddnn}{\delta_{j,\{n,n'\}}^{(4)}}
\nc{\dddnn}{\delta_{j,\{n,n'\}}^{(8)}}

\definecolor{pink}{rgb}{1,.65,.65}

\begin{document}

\topmargin -5mm
\oddsidemargin 5mm

\setcounter{page}{1}

\vspace{8mm}
\begin{center}
{\huge {\bf ${\cal W}$-Extended Logarithmic Minimal Models}}

\vspace{10mm}
 {\LARGE J{\o}rgen Rasmussen}
\\[.3cm]
 {\em Department of Mathematics and Statistics, University of Melbourne}\\
 {\em Parkville, Victoria 3010, Australia}
\\[.4cm]
 {\tt j.rasmussen@ms.unimelb.edu.au}

\end{center}

\vspace{8mm}
\centerline{{\bf{Abstract}}}
\vskip.4cm
\noindent
We consider the continuum scaling limit of the infinite 
series of Yang-Baxter integrable logarithmic minimal models ${\cal LM}(p,p')$ 
as `rational' logarithmic conformal field theories with extended ${\cal W}$ symmetry. 
The representation content is found to consist of
$6pp'-2p-2p'$ ${\cal W}$-indecomposable representations of which $2p+2p'-2$ 
are of rank 1, $4pp'-2p-2p'$ are of rank 2, while the remaining $2(p-1)(p'-1)$ are of rank 3. 
We identify these representations with suitable limits of Yang-Baxter integrable boundary 
conditions on the lattice. 
The ${\cal W}$-indecomposable rank-1 representations are all ${\cal W}$-irreducible while 
we present a conjecture for the embedding patterns of the ${\cal W}$-indecomposable
rank-2 and -3 representations.
The associated ${\cal W}$-extended characters are all given explicitly and
decompose as finite non-negative sums of ${\cal W}$-irreducible characters.
The latter correspond to ${\cal W}$-irreducible subfactors and we find
that there are $2pp'+(p-1)(p'-1)/2$ of them. We present fermionic character expressions for some of 
the rank-2 and all of the rank-3 ${\cal W}$-indecomposable representations. To distinguish between
inequivalent ${\cal W}$-indecomposable representations of identical characters,
we introduce `refined' characters carrying information also about the Jordan-cell content 
of a representation. Using a lattice implementation of fusion on a strip, 
we study the fusion rules for the ${\cal W}$-indecomposable representations and find that 
they generate a closed fusion algebra, albeit one without identity for $p>1$.  
We present the complete set of fusion rules and
interpret the closure of this fusion algebra as confirmation of the proposed extended symmetry. 
Finally, $2pp'$ of the ${\cal W}$-indecomposable representations are in fact
${\cal W}$-projective representations and they generate a closed fusion subalgebra.
\renewcommand{\thefootnote}{\arabic{footnote}}
\setcounter{footnote}{0}

\section{Introduction}

We consider the infinite series of Yang-Baxter integrable logarithmic minimal models 
${\cal LM}(p,p')$~\cite{PRZ0607}. These are examples of two-dimensional lattice systems 
whose continuum scaling limits~\cite{Cardy87} give rise to conformal field theories (CFTs).
Our lattice approach to studying these CFTs is predicated on the supposition that, 
in the continuum scaling limit, a transfer matrix with prescribed boundary conditions gives rise to 
a representation of the Virasoro algebra. Different boundary conditions naturally lead to 
different representations. We further assume that, if in addition, the boundary conditions 
respect the symmetry of a larger conformal algebra ${\cal W}$~\cite{Zam85,BS93}, 
then the continuum scaling 
limit of the transfer matrix will yield a representation of the extended algebra ${\cal W}$.

A central question of much current interest~\cite{Flohr96,GK9606,FG05,GR06} is whether an 
extended symmetry algebra ${\cal W}$ exists for logarithmic 
CFTs~\cite{Gur93,Flo03,Gab03,Kaw03} like the logarithmic minimal models. 
Such a symmetry should allow the countably {\em infinite} number of Virasoro representations 
to be reorganized into a {\em finite} number of extended ${\cal W}$-representations which close 
under fusion. In the case of the logarithmic minimal models ${\cal LM}(1,p')$, the existence of 
such an extended ${\cal W}$-symmetry and the associated fusion rules are by now 
well established~\cite{GK9606,FHST03,FGST05,GR07,GTipunin07,PRR0803}. 
By stark contrast, although there are strong indications~\cite{FGST06a,FGST06b} that there exists a 
${\cal W}_{p,p'}$ symmetry algebra for general augmented minimal models, very little is known 
about the ${\cal W}$-extended fusion rules for the ${\cal LM}(p,p')$ models with $p\ge 2$.
This situation was partly resolved in our recent paper~\cite{RP0804}. 
There we used a lattice approach on a strip, generalizing the 
approach of~\cite{PRR0803}, to obtain fusion rules of critical percolation ${\cal LM}(2,3)$ in the 
extended picture. 

In~\cite{PRR0803}, it was shown that symplectic fermions~\cite{Kausch95,Kausch00}
is just critical dense polymers 
${\cal LM}(1,2)$~\cite{Gennes,Cloizeaux,Saleur86,Saleur87,Duplantier,PR0610} 
viewed in the extended picture. 
Likewise in the case of critical percolation~\cite{RP0804}, the extended picture is described
by the {\em same} lattice model as the Virasoro picture~\cite{PRZ0607,RP0706}. 
It is nevertheless useful to distinguish between the two pictures by denoting
the extended picture by ${\cal WLM}(2,3)$ and reserve the notation ${\cal LM}(2,3)$ for 
critical percolation in the non-extended Virasoro picture. A similar distinction applies to the
entire infinite series of logarithmic minimal models. 
Their extended pictures are thus denoted by ${\cal WLM}(p,p')$ and are the topic of the present work.
The ${\cal W}$-extended fusion rules we obtain for these models
are based on the {\em fundamental\/} fusion algebra in the Virasoro 
picture~\cite{RP0706,RP0707} which is a subset of the {\em full\/} fusion algebra. 
The latter remains to be determined and may eventually yield a larger ${\cal W}$-extended
fusion algebra than the one presented here. 

The layout of this paper is as follows. In Section~\ref{SectionLM}, we review the logarithmic 
minimal model ${\cal LM}(p,p')$ and its fusion algebra~\cite{PRZ0607,RP0707}. 
In Section~\ref{SectionWLM}, we first summarize the 
representation content of ${\cal WLM}(p,p')$ consisting
of $6pp'-2p-2p'$ ${\cal W}$-indecomposable representations of which $2p+2p'-2$ 
are of rank 1, $4pp'-2p-2p'$ are of rank 2, while the remaining $2(p-1)(p'-1)$ are of rank 3. 
The ${\cal W}$-indecomposable rank-1 representations are all ${\cal W}$-irreducible while 
we present a conjecture for the embedding patterns of the ${\cal W}$-indecomposable
rank-2 and -3 representations. The associated ${\cal W}$-extended characters
decompose as finite non-negative sums of ${\cal W}$-irreducible characters.
The latter correspond to ${\cal W}$-irreducible subfactors and we find
that there are $2pp'+(p-1)(p'-1)/2$ of them. These are all identified.
To distinguish between
inequivalent ${\cal W}$-indecomposable representations of identical characters,
we introduce `refined' characters carrying information also about the Jordan-cell content 
of a representation. We then present
the fusion rules demonstrating closure of the associative and commutative
fusion algebra of the ${\cal W}$-indecomposable representations. 
We conclude this section with a discussion of the ${\cal W}$-projective representations
and find that there are $2pp'$ of them and that they generate a closed fusion
subalgebra. In Section~\ref{SectionLattice}, we identify the ${\cal W}$-extended representations 
with suitable limits of Yang-Baxter integrable boundary conditions on the lattice and give details of 
their construction and properties. 
In particular, we explain how fusion is implemented on the lattice and analyze the ensuing 
closed fusion algebra. For $p>1$, this fusion algebra does not contain an identity.
We conclude with a short discussion in Section~\ref{SectionDiscussion}.

\subsection*{Notation}
\vskip.1cm 
For $n,m\in\mathbb{Z}$,
\be
 \mathbb{Z}_{n,m}\ =\ \mathbb{Z}\cap[n,m]
\ee 
denotes the set of integers from $n$ to $m$, both included.
Certain properties of integers modulo 2 are written
\be
 \eps(n)\ =\ \frac{1-(-1)^n}{2},\qquad\quad
   n\cdot m\ =\ \frac{3-(-1)^{n+m}}{2},\qquad\quad n,m\in\mathbb{Z}
\label{eps}
\ee
where the dot product is seen to be associative.
An $n$-fold fusion of the Virasoro representation $A$ with itself is abbreviated
\be
 A^{\otimes n}\ =\ \underbrace{A\otimes A\otimes\ldots\otimes A}_n
\ee
By a direct sum of representations $A_n$
with unspecified lower summation bound, we mean a direct sum in steps of 2
whose lower bound is given by the parity $\eps(N)$ of the upper bound, that is,
\be
 \bigoplus_{n}^{N}A_n\ =\ \bigoplus_{n=\eps(N),\ \!{\rm by}\ \!2}^{N}A_n,\qquad\quad N\in\mathbb{Z}
\label{sum2}
\ee
This direct sum vanishes for negative $N$.

\section{Logarithmic minimal model ${\cal LM}(p,p')$}
\label{SectionLM}

A logarithmic minimal model ${\cal LM}(p,p')$ is defined \cite{PRZ0607} for every coprime pair of
positive integers $p<p'$.
The model ${\cal LM}(p,p')$ has central charge
\be
 c\ =\  1-6\frac{(p'-p)^2}{pp'}
\label{c}
\ee
and conformal weights
\be
 \D_{r,s}\ =\ \frac{(rp'-sp)^{2}-(p'-p)^2}{4pp'},\hspace{1.2cm} r,s\in\mathbb{N}
\label{D}
\ee
The fundamental fusion algebra $\big\langle(2,1),(1,2)\big\rangle_{p,p'}$ \cite{RP0706,RP0707}
of the logarithmic
minimal model ${\cal LM}(p,p')$ is generated by the two fundamental Kac representations
$(2,1)$ and $(1,2)$ and contains a countably infinite number of inequivalent, indecomposable 
representations of rank 1, 2 or 3. 
For $r,s\in\mathbb{N}$, the character of the Kac representation $(r,s)$ is
\be
 \chit_{r,s}(q)\ =\ \frac{q^{\frac{1-c}{24}+\D_{r,s}}}{\eta(q)}\big(1-q^{rs}\big)
  \ =\ \frac{1}{\eta(q)}\big(q^{(rp'-sp)^2/4pp'}-q^{(rp'+sp)^2/4pp'}\big)
\label{chikac}
\ee
where the Dedekind eta function is given by
\be
  \eta(q)\ =\ q^{\frac{1}{24}} \prod_{n=1}^\infty (1-q^n)
\label{eta}
\ee
Such a representation is of rank 1 and is irreducible if $r\in\mathbb{Z}_{1,p}$ and $s\in p'\mathbb{N}$
or if $r\in p\mathbb{N}$ and $s\in\mathbb{Z}_{1,p'}$. It is a reducible yet indecomposable
representation if $r\in\mathbb{Z}_{1,p-1}$ and $s\in\mathbb{Z}_{1,p'-1}$,
while it is a fully reducible representation if $r\in p\mathbb{N}$ and $s\in p'\mathbb{N}$ where
\be
 (kp,k'p')\ =\ (k'p,kp')\ =\ \bigoplus_{j=|k-k'|+1,\ \!{\rm by}\ \!2}^{k+k'-1}(jp,p')\ =\ 
   \bigoplus_{j=|k-k'|+1,\ \!{\rm by}\ \!2}^{k+k'-1}(p,jp')
\label{kpkp}
\ee
These are the only Kac representations appearing in the fundamental fusion algebra.
The characters of the reducible yet indecomposable Kac representations just mentioned
can be written as sums of two irreducible Virasoro characters
\be
 \chit_{r,s}(q)\ =\ \ch_{r,s}(q)+\ch_{2p-r,s}(q)\ =\ \ch_{r,s}(q)+\ch_{r,2p'-s}(q),\qquad
   r\in\mathbb{Z}_{1,p-1},\quad s\in\mathbb{Z}_{1,p'-1}
\ee
In general and with $a\in\mathbb{Z}_{1,p-1}$, $b\in\mathbb{Z}_{1,p'-1}$ and $k\in\mathbb{N}-1$, 
the irreducible Virasoro characters read \cite{FSZ}
\bea
 {\rm ch}_{a+kp,b}(q)&=&K_{2pp',(a+kp)p'-bp;k}(q)-K_{2pp',(a+kp)p'+bp;k}(q)\nn
 {\rm ch}_{a+(k+1)p,p'}(q)&=&
  \frac{1}{\eta(q)}\big(q^{(kp+a)^2p'/4p}-q^{((k+2)p-a)^2p'/4p}\big) \nn
 {\rm ch}_{(k+1)p,b}(q)&=&
  \frac{1}{\eta(q)}\big(q^{((k+1)p'-b)^2p/4p'}-q^{((k+1)p'+b)^2p/4p'}\big) \nn
 {\rm ch}_{(k+1)p,p'}(q)&=&
  \frac{1}{\eta(q)}\big(q^{k^2pp'/4}-q^{(k+2)^2pp'/4}\big)
\label{laq}
\eea
where $K_{n,\nu;k}(q)$ is defined as
\be
 K_{n,\nu;k}(q)\ =\ \frac{1}{\eta(q)}\sum_{j\in\mathbb{Z}\setminus\mathbb{Z}_{1,k}}q^{(\nu-jn)^2/2n}
\label{Kk}
\ee
For $r\in\mathbb{Z}_{1,p}$, $s\in\mathbb{Z}_{1,p'}$, $a\in\mathbb{Z}_{1,p-1}$, 
$b\in\mathbb{Z}_{1,p'-1}$ and $k\in\mathbb{N}$, 
the representations denoted by $\R_{kp,s}^{a,0}$ and $\R_{r,kp'}^{0,b}$ are indecomposable
representations of rank 2, while $\R_{kp,p'}^{a,b}\equiv\R_{p,kp'}^{a,b}$ 
is an indecomposable representation of rank 3. Their characters read
\bea
 \chit[\R_{kp,s}^{a,0}](q)&=&
    \big(1-\delta_{k,1}\delta_{s,p'}\big)\ch_{kp-a,s}(q)+2\ch_{kp+a,s}(q)+\ch_{(k+2)p-a,s}(q)\nn
 \chit[\R_{r,kp'}^{0,b}](q)&=&
    \big(1-\delta_{k,1}\delta_{r,p}\big)\ch_{r,kp'-b}(q)+2\ch_{r,kp'+b}(q)+\ch_{r,(k+2)p'-b}(q)\nn
 \chit[\R_{kp,p'}^{a,b}](q)&=&\big(1-\delta_{k,1}\big)\ch_{(k-1)p-a,b}(q)+2\ch_{(k-1)p+a,b}(q)
   +2\big(1-\delta_{k,1}\big)\ch_{kp-a,p'-b}(q)\nn
   &&+\ 4\ch_{kp+a,p'-b}(q)+\big(2-\delta_{k,1}\big)\ch_{(k+1)p-a,b}(q)
  +2\ch_{(k+1)p+a,b}(q)\nn
   &&+\ 2\ch_{(k+2)p-a,p'-b}(q)+\ch_{(k+3)p-a,b}(q)
\label{chiR}
\eea
Indecomposable representations of rank 3 appear for $p>1$ only.
For $\al\in\mathbb{Z}_{0,p-1}$, $\beta\in\mathbb{Z}_{0,p'-1}$ and $k,k'\in\mathbb{N}$, 
a decomposition similar to (\ref{kpkp}) also applies to the higher-rank {\em decomposable}
representations $\R_{kp,k'p'}^{\al,\beta}$ as we have 
\be
 \R_{kp,k'p'}^{\al,\beta}\ =\ \R_{k'p,kp'}^{\al,\beta}
   \ =\ \bigoplus_{j=|k-k'|+1,\ \!{\rm by}\ \!2}^{k+k'-1}\R_{jp,p'}^{\al,\beta}
   \ =\ \bigoplus_{j=|k-k'|+1,\ \!{\rm by}\ \!2}^{k+k'-1}\R_{p,jp'}^{\al,\beta}
\ee
Here we have introduced the convenient notation
\be
 \R_{r,s}^{0,0}\ \equiv\ (r,s),\qquad\quad r,s\in\mathbb{N}
\label{R00}
\ee
In the following, we will also use
\be
 \R_{0,s}^{\al,\beta}\ \equiv\ \R_{r,0}^{\al,\beta}\ \equiv\ 0,\qquad\quad 
   \al\in\mathbb{Z}_{0,p-1},\quad
   \beta\in\mathbb{Z}_{0,p'-1},\quad r,s\in\mathbb{N}
\ee

Fusion in the fundamental fusion algebra $\big\langle(2,1),(1,2)\big\rangle_{p,p'}$
decomposes into `horizontal' and `vertical' components. With 
$\al\in\mathbb{Z}_{0,p-1}$, $\beta\in\mathbb{Z}_{0,p'-1}$ and $k\in\mathbb{N}$, we thus have 
\be
 \R_{p,kp'}^{\al,\beta}\ =\ \R_{p,1}^{\al,0}\otimes\R_{1,kp'}^{0,\beta}
   \ =\ \R_{kp,1}^{\al,0}\otimes\R_{1,p'}^{0,\beta}
\label{decomp}
\ee
The Kac representation $(1,1)$ is the identity of the fundamental fusion algebra.
For $p>1$, this is a reducible yet indecomposable representation, while for $p=1$, it
is an irreducible representation.

Finally, for later reference, we list the horizontal fusions
\be
 (r,1)\otimes(r',1)\ =\ \bigoplus_{j=|r-r'|+1,\ \!{\rm by}\ \!2}^{p-|p-r-r'|-1}(j,1)
   \oplus\bigoplus_{\al}^{r+r'-p-1}\R_{p,1}^{\al,0},
      \qquad\quad r,r'\in\mathbb{Z}_{1,p}
\label{r1r1}
\ee
and
\bea
 (kp,1)\otimes(k'p,1)&=&\!\!\bigoplus_{j=|k-k'|+1,\ \!{\rm by}\ \!2}^{k+k'-1}
   \Big\{\bigoplus_{\al}^{p-1}\R_{jp,1}^{\al,0}\Big\}\nn
 \R_{kp,1}^{a,0}\otimes(k'p,1)&=&\!\!\bigoplus_{j=|k-k'|,\ \!{\rm by}\ \!2}^{k+k'}
  \dkk \Big\{\bigoplus_{\al}^{a-1}\R_{jp,1}^{\al,0}\Big\}
   \oplus\!\bigoplus_{j=|k-k'|+1,\ \!{\rm by}\ \!2}^{k+k'-1}
  \!\Big\{\bigoplus_{\al}^{p-a-1}2\R_{jp,1}^{\al,0}\Big\}
\label{kpkpRkpkp}
\eea
recalling the meaning of a direct sum with unspecified lower bound (\ref{sum2}),
and the horizontal triple fusions
\bea
 (p,1)^{\otimes3}&=&\bigoplus_{\al=0,\ \!{\rm by}\ \!2}^{p-2}(p-\al)
   \Big\{\R_{p,1}^{\al,0}\oplus\frac{1}{2}\R_{2p,1}^{\al,0}\Big\},\qquad\qquad\qquad\qquad
      \quad p\ \mathrm{even}\nn
 (p,1)^{\otimes3}&=&\bigoplus_{\al=0,\ \!{\rm by}\ \!2}^{p-1}(p-\al)\R_{p,1}^{\al,0}
  \oplus\!\!\bigoplus_{\al=1,\ \!{\rm by}\ \!2}^{p-2}\frac{1}{2}(p-\al)\R_{2p,1}^{\al,0},
    \qquad\qquad p\ \mathrm{odd}
\label{ppp}
\eea
The Kronecker delta function combination \cite{RP0706} appearing in (\ref{kpkpRkpkp}) is defined by
\be
 \dkk\ =\ 2-\delta_{j,|k-k'|}-\delta_{j,k+k'}
\label{dkk}
\ee
Despite the factors of $1/2$ in the decompositions (\ref{ppp}), the multiplicities are all integer.
Similar expressions for vertical fusions naturally apply.
We refer to \cite{RP0707} for further details on the fundamental fusion algebra
of ${\cal LM}(p,p')$.

\section{${\cal W}$-extended logarithmic minimal model ${\cal WLM}(p,p')$}
\label{SectionWLM}

In this section, we summarize our findings in the extended picture 
${\cal WLM}(p,p')$ for the representation content and the associated embedding patterns, 
(refined) characters and closed fusion algebra. 
Unless otherwise specified, we let
\be
 \kappa,\kappa'\in\mathbb{Z}_{1,2},\quad r\in\mathbb{Z}_{1,p},\quad
  s\in\mathbb{Z}_{1,p'},\quad a,a'\in\mathbb{Z}_{1,p-1},\quad b,b'\in\mathbb{Z}_{1,p'-1},
  \quad\al\in\mathbb{Z}_{0,p-1},\quad\beta\in\mathbb{Z}_{0,p'-1}
\ee
and $k,k',n\in\mathbb{N}$.

\subsection{Representation content}

We have the $2(p+p'-1)$ ${\cal W}$-indecomposable rank-1 representations
\be
 \big\{\ketw{\kappa p,s},\ketw{r,\kappa p'}\big\}
 \hspace{1.2cm}\mathrm{subject\ to}\ \ \ \ketw{p,\kappa p'}\equiv\ketw{\kappa p,p'}
\label{r1}
\ee
where $\ketw{p,p'}$ is listed twice,
the $2\big((p-1)p'+p(p'-1)\big)$ ${\cal W}$-indecomposable rank-2 representations
\be
 \big\{\ketw{\R_{\kappa p,s}^{a,0}}, \ketw{\R_{r,\kappa p'}^{0,b}}\big\}
\label{r2}
\ee
and the $2(p-1)(p'-1)$ ${\cal W}$-indecomposable rank-3 representations
\be
 \big\{\ketw{\R_{\kappa p,\kappa' p'}^{a,b}}\big\}
 \hspace{1.2cm}\mathrm{subject\ to}\ \ \ 
  \ketw{\R_{p,2p'}^{a,b}}\equiv\ketw{\R_{2p,p'}^{a,b}}\quad
   \mathrm{and}\quad \ketw{\R_{2p,2p'}^{a,b}}\equiv\ketw{\R_{p,p'}^{a,b}}
\label{r3}
\ee
Here we are asserting that these ${\cal W}$-representations are indeed 
${\cal W}$-{\rm indecomposable}. We furthermore believe that the ${\cal W}$-indecomposable
representations (\ref{r1}) are ${\cal W}$-{\em irreducible}.
Compactly, the various ${\cal W}$-indecomposable representations satisfy 
\be
 \ketw{\R_{(\kappa\cdot\kappa')p,p'}^{\al,\beta}}
  \ \equiv\ \ketw{\R_{\kappa p,\kappa' p'}^{\al,\beta}}
  \ \equiv\ \ketw{\R_{p,(\kappa\cdot\kappa')p'}^{\al,\beta}}
\label{identi}
\ee
where we have extended our notation to include $\ketw{\R_{\kappa p,\kappa'p'}^{\al,\beta}}$ for all
$\kappa,\kappa'\in\mathbb{Z}_{1,2}$.
The numbers of rank-1 -2 and -3 ${\cal W}$-indecomposable representations are thus
\be
 N_1(p,p')\ =\ 2(p+p'-1),\qquad
 N_2(p,p')\ =\ 2(2pp'-p-p'),\qquad
 N_3(p,p')\ =\ 2(p-1)(p'-1)
\ee
respectively. The total number of ${\cal W}$-indecomposable representations is therefore given by
\be
 N_{\mathrm{ind}}(p,p')\ =\ 6pp'-2(p+p')
\label{Ntotal}
\ee
In the case of ${\cal LM}(1,p')$, we recover the well-known numbers
\be
 N_1(1,p')\ =\ 2p',\qquad
 N_2(1,p')\ =\ 2(p'-1),\qquad
 N_3(1,p')\ =\ 0,\qquad
 N_{\mathrm{ind}}(1,p')\ =\ 4p'-2
\ee
The ${\cal W}$-extended logarithmic minimal models ${\cal WLM}(1,p')$ are discussed in 
\cite{GK9606,FHST03,FGST05,GR07,GTipunin07,PRR0803}
while ${\cal W}$-extended critical percolation ${\cal WLM}(2,3)$ is discussed in \cite{RP0804}.
In the latter case, the numbers are
\be
 N_1(2,3)\ =\ 8,\qquad
 N_2(2,3)\ =\ 14,\qquad
 N_3(2,3)\ =\ 4,\qquad
 N_{\mathrm{ind}}(2,3)\ =\ 26
\ee

In terms of Virasoro-indecomposable representations, the ${\cal W}$-indecomposable rank-1 representations decompose as
\bea
 \ketw{\kappa p,s}&=&\bigoplus_{k\in\mathbb{N}}(2k-2+\kappa)((2k-2+\kappa)p,s)\nn
 \ketw{r,\kappa p'}&=&\bigoplus_{k\in\mathbb{N}}(2k-2+\kappa)(r,(2k-2+\kappa)p')
\label{r1Vir}
\eea
where the two expressions for $\ketw{p,p'}$ agree and where the identity
$\ketw{p,2p'}\equiv\ketw{2p,p'}$ is verified explicitly.
Similarly, the ${\cal W}$-indecomposable rank-2 representations decompose as
\be
 \ketw{\R_{\kappa p,s}^{a,0}}\ =\ 
   \bigoplus_{k\in\mathbb{N}}(2k-2+\kappa)\R_{(2k-2+\kappa)p,s}^{a,0},\qquad\quad
 \ketw{\R_{r,\kappa p'}^{0,b}}\ =\ \bigoplus_{k\in\mathbb{N}}(2k-2+\kappa)\R_{r,(2k-2+\kappa)p'}^{0,b}
\label{r2Vir}
\ee
while the ${\cal W}$-indecomposable rank-3 representations decompose as
\be
 \ketw{\R_{\kappa p,p'}^{a,b}}  
   \ =\ \bigoplus_{k\in\mathbb{N}}(2k-2+\kappa)\R_{p,(2k-2+\kappa)p'}^{a,b}
   \ =\ \bigoplus_{k\in\mathbb{N}}(2k-2+\kappa)\R_{(2k-2+\kappa)p,p'}^{a,b}
\label{r3Vir}
\ee

\subsection{${\cal W}$-extended characters}

The characters of the ${\cal W}$-indecomposable rank-1 representations read
\bea
 \chih_{\kappa p,s}(q)&=&\sum_{k\in\mathbb{N}}(2k-2+\kappa)\ch_{(2k-2+\kappa)p,s}(q)
   \ =\ \frac{1}{\eta(q)}\sum_{k\in\mathbb{Z}}(2k-2+\kappa)q^{((2k-2+\kappa)p'-s)^2p/4p'}\nn
 \chih_{r,\kappa p'}(q)&=&\sum_{k\in\mathbb{N}}(2k-2+\kappa)\ch_{r,(2k-2+\kappa)p'}(q)
   \ =\ \frac{1}{\eta(q)}\sum_{k\in\mathbb{Z}}(2k-2+\kappa)q^{((2k-2+\kappa)p-r)^2p'/4p} 
\eea
where it is recalled that $\ketw{2p,p'}\equiv\ketw{p,2p'}$.
The characters of the ${\cal W}$-indecomposable rank-2 representations read 
\bea
 \chit\big[\ketw{\R_{\kappa p,s}^{a,0}}\big](q)
   &=&\delta_{\kappa,1}\big(1-\delta_{s,p'}\big)\ch_{p-a,s}(q)
  +2\sum_{k\in\mathbb{N}}(2k+1-\kappa)\ch_{(2k+2-\kappa)p-a,s}(q)\nn
   &&\hspace{1.2cm}
      +\ 2\sum_{k\in\mathbb{N}}(2k-2+\kappa)\ch_{(2k-2+\kappa)p+a,s}(q)\nn
 \chit\big[\ketw{\R_{r,\kappa p'}^{0,b}}\big](q)
   &=&\delta_{\kappa,1}\big(1-\delta_{r,p}\big)\ch_{r,p'-b}(q)
  +2\sum_{k\in\mathbb{N}}(2k+1-\kappa)\ch_{r,(2k+2-\kappa)p'-b}(q)\nn
   &&\hspace{1.2cm}
      +\ 2\sum_{k\in\mathbb{N}}(2k-2+\kappa)\ch_{r,(2k-2+\kappa)p'+b}(q)
\eea
that is,
\bea
  \chit\big[\ketw{\R_{\kappa p,b}^{a,0}}\big](q)
   &=&\frac{1}{\eta(q)}\sum_{k\in\mathbb{Z}}(2k-2+\kappa)\Big(
     q^{(ap'-bp+(2k-2+\kappa)pp')^2/4pp'}-q^{(ap'+bp+(2k-2+\kappa)pp')^2/4pp'}\Big)\nn
  \chit\big[\ketw{\R_{\kappa p,p'}^{a,0}}\big](q)
   &=& \frac{2}{\eta(q)}\sum_{k\in\mathbb{Z}}q^{(a+(2k-1+\kappa)p)^2p'/4p}\nn
  \chit\big[\ketw{\R_{a,\kappa p'}^{0,b}}\big](q)
   &=&\frac{1}{\eta(q)}\sum_{k\in\mathbb{Z}}(2k-2+\kappa)\Big(
     q^{(-ap'+bp+(2k-2+\kappa)pp')^2/4pp'}-q^{(ap'+bp+(2k-2+\kappa)pp')^2/4pp'}\Big)\nn
  \chit\big[\ketw{\R_{p,\kappa p'}^{0,b}}\big](q)
   &=& \frac{2}{\eta(q)}\sum_{k\in\mathbb{Z}}q^{(b+(2k-1+\kappa)p')^2p/4p'}
\label{r2q}
\eea
We note the character identities
\be
 \chit\big[\ketw{\R_{p,p'}^{a,0}}\big](q)\ =\ \chit\big[\ketw{\R_{2p,p'}^{p-a,0}}\big](q),\qquad\quad
 \chit\big[\ketw{\R_{p,p'}^{0,b}}\big](q)\ =\ \chit\big[\ketw{\R_{p,2p'}^{0,p'-b}}\big](q) 
\label{r2chid}
\ee
and the character relations
\bea
 \chit\big[\ketw{\R_{p,b}^{a,0}}\big](q)
   &=&\ch_{p-a,b}(q)+\chit\big[\ketw{\R_{2p,b}^{p-a,0}}\big](q)\nn
 \chit\big[\ketw{\R_{a,p'}^{0,b}}\big](q)&=&\ch_{a,p'-b}(q)+\chit\big[\ketw{\R_{a,2p'}^{0,p'-b}}\big](q) 
\eea
and
\be 
 \chit\big[\ketw{\R_{\kappa p,b}^{a,0}}\big](q)+\chit\big[\ketw{\R_{\kappa p,p'-b}^{p-a,0}}\big](q)
  \ =\  \chit\big[\ketw{\R_{a,\kappa p'}^{0,b}}\big](q)+\chit\big[\ketw{\R_{p-a,\kappa p'}^{0,p'-b}}\big](q)
\ee
The characters of the ${\cal W}$-indecomposable rank-3 representations read 
\bea
 \chit[\ketw{\R_{\kappa p,p'}^{a,b}}](q)
   &=&2\delta_{\kappa,1}\ch_{a,b}(q)+2\delta_{\kappa,2}\ch_{p-a,b}(q)\nn
   &+&4\sum_{k\in\mathbb{N}}(2k-2+\kappa)\big(\ch_{(2k-2+\kappa)p+a,p'-b}(q)
       +\ch_{(2k-2+\kappa)p+p-a,b}(q)\big)\nn
   &+&4\sum_{k\in\mathbb{N}}(2k+1-\kappa)\big(\ch_{(2k+1-\kappa)p+a,b}(q)
      +\ch_{a,(2k+1-\kappa)p'+b}(q)\big)\nn
  &=&\frac{2}{\eta(q)}\sum_{k\in\mathbb{Z}}\Big(q^{(ap'-bp+(2k+1-\kappa)pp')^2/4pp'}
    +q^{(ap'+bp+(2k+1-\kappa)pp')^2/4pp'}\Big)
\label{r3q}
\eea
and satisfy
\be
 \chit\big[\ketw{\R_{(3-\kappa) p,p'}^{a,b}}\big](q)
  \ =\ \chit\big[\ketw{\R_{\kappa p,p'}^{p-a,b}}\big](q)
  \ =\ \chit\big[\ketw{\R_{\kappa p,p'}^{a,p'-b}}\big](q)
\label{r3chid}
\ee
As we will discuss below, the rank-2 and -3 representations listed in (\ref{r2})
and (\ref{r3}) all have distinct Jordan-cell and general embedding
structures, despite the character identities (\ref{r2chid}) and (\ref{r3chid}). 

It is pointed out that some of the character expressions above are {\em fermionic},
namely $\chit\big[\ketw{\R_{\kappa p,p'}^{a,0}}\big](q)$ and 
$\chit\big[\ketw{\R_{p,\kappa p'}^{0,b}}\big](q)$ in (\ref{r2q}) and 
$\chit[\ketw{\R_{\kappa p,p'}^{a,b}}](q)$ in (\ref{r3q}).
Although of great interest, it is beyond the scope of the present paper to work out 
fermionic character expressions for the other ${\cal W}$-indecomposable representations.

\subsubsection{Irreducible subfactors}
\label{SectionIrrSub}

It is recalled that Virasoro-irreducible characters are denoted by $\ch_{\rho,\sigma}(q)$ 
where $\rho,\sigma\in\mathbb{N}$ as we reserve the notation
$\chit_{\rho,\sigma}(q)$ for the characters of the (in general reducible) 
Kac representations $(\rho,\sigma)$. Only if the Kac representation happens
to be Virasoro-irreducible, cf. the discussion following (\ref{eta}), do we use both notations.
In the ${\cal W}$-extended picture, on the other hand, we will denote the character of 
a ${\cal W}$-irreducible representation of conformal weight $\D_{\rho,\sigma}$ simply by 
$\chih_{\rho,\sigma}(q)$.

Here we assert that, in addition to the $2(p+p'-1)$ ${\cal W}$-irreducible rank-1
representations listed in (\ref{r1}), there are $\frac{5}{2}(p-1)(p'-1)$ ${\cal W}$-irreducible rank-1
representations appearing as {\em subfactors} of the ${\cal W}$-indecomposable
rank-2 and -3 representations. 
This brings the total number of ${\cal W}$-irreducible characters to 
\be
 N_{\mathrm{irr}}(p,p')\ =\ N_1+\frac{5}{2}(p-1)(p'-1)
   \ =\ 2pp' +\frac{1}{2}(p-1)(p'-1)
\ee
$\frac{1}{2}(p-1)(p'-1)$ of these new ${\cal W}$-irreducible representations 
simply correspond to Virasoro-irreducible representations and have characters given by
\be
  \chih_{a,b}(q)\ =\ \chih_{p-a,p'-b}(q)\ =\ \ch_{a,b}(q)
   \ =\ \frac{1}{\eta(q)}\sum_{k\in\mathbb{Z}}
    \Big(q^{(ap'-bp+2kpp')^2/4pp'}-q^{(ap'+bp+2kpp')^2/4pp'}\Big)
\label{chihab}
\ee
The remaining $2(p-1)(p'-1)$ new ${\cal W}$-irreducible representations have characters
\bea
 \!\!\!\chih_{\kappa p+a,b}(q)&=&\sum_{k\in\mathbb{N}}(2k-2+\kappa)\ch_{(2k-2+\kappa)p+a,b}(q) \nn
  &=& \frac{1}{\eta(q)}\sum_{k\in\mathbb{Z}}k(k-1+\kappa)\Big(
    q^{(ap'-bp+(2k-2+\kappa)pp')^2/4pp'}-q^{(ap'+bp+(2k-2+\kappa)pp')^2/4pp'}\Big) 
\label{chih}
\eea
satisfying
\be
    \chih_{\kappa p+a,p'-b}(q)\ =\ \chih_{p-a,\kappa p'+b}(q)
\ee
We can now express the characters of the higher-rank ${\cal W}$-indecomposable representations
in terms of ${\cal W}$-irreducible characters, and we find that the rank-2 and -3 characters 
enjoy the decompositions
\bea
 \chit\big[\ketw{\R_{\kappa p,s}^{a,0}}\big](q)
   &=& \delta_{\kappa,1}\big(1-\delta_{s,p'}\big)\chih_{p-a,s}(q)+2\chih_{(4-\kappa)p-a,s}(q)
    +2\chih_{\kappa p+a,s}(q)\nn
 \chit\big[\ketw{\R_{r,\kappa p'}^{0,b}}\big](q)&=&
    \delta_{\kappa,1}\big(1-\delta_{r,p}\big)\chih_{r,p'-b}(q)+2\chih_{r,(4-\kappa)p'-b}(q)
    +2\chih_{r,\kappa p'+b}(q)
\eea
and
\bea
 \chit\big[\ketw{\R_{\kappa p,p'}^{a,b}}\big](q)
   &=& 2\delta_{\kappa,1}\chih_{a,b}(q)+2\delta_{\kappa,2}\chih_{p-a,b}(q)
    +4\chih_{\kappa p+a,p'-b}(q)\nn
   &+&4\chih_{\kappa p+p-a,b}(q)+4\chih_{(3-\kappa)p+a,b}(q)
    +4\chih_{a,(3-\kappa)p'+b}(q)
\eea

\subsubsection{Theta forms}

The characters of the $N_{\mathrm{irr}}(p,p')$ ${\cal W}$-irreducible representations agree with 
those of \cite{FGST06b}.Ê
In particular, they admit the expressions given there in terms of theta functions
\be
  \theta_{\ell,k}(q,z)\ =\ \sum_{j\in\mathbb{Z}+\frac{\ell}{2k}} 
    q^{kj^2} z^{k j},\qquad |q|<1,\quad z\in\mathbb{C},\quad k\in\mathbb{N},\quad \ell\in\mathbb{Z}
\ee
and so-called theta-constants
\be
  \theta_{\ell,k}(q)\ =\ \theta_{\ell,k}(q,1),\quad \theta_{\ell,k}^{(m)}(q)
    \ =\ \bigg(z\frac{\partial}{\partial z}\bigg)^m\theta_{\ell,k}(q,z)\bigg|_{z=1},\qquad m\in\mathbb{N}
\ee
Introducing the abbreviations
\be
  \theta_\ell(q)\ =\ \theta_{\ell,pp'}(q),\qquad \theta'_\ell(q)\ =\ \theta_{\ell,pp'}^{(1)}(q),\qquad 
    \theta''_\ell(q)\ =\ \theta^{(2)}_{\ell,pp'}(q)
\ee
the theta forms are
\bea
 \chih_{a,b}(q)&=&\frac{1}{\eta(q)}(\theta_{sp-rp'}(q)-\theta_{sp+rp'}(q))
\label{tfab}
\\
 \chih^+_{r,s}(q)&=&\frac{1}{(pp')^2\eta(q)}\bigg(\theta''_{sp+rp'}(q)
   -\theta''_{sp-rp'}(q)-(sp+rp')\theta'_{sp+rp'}(q)+(sp-rp')\theta'_{sp-rp'}(q)\nonumber\\
 &&\mbox{}\qquad\qquad\qquad
   +\frac{(sp+rp')^2}{4}\,\theta_{sp+rp'}(q)-\frac{(sp-rp')^2}{4}\,\theta_{sp-rp'}(q)\bigg)\\
  \chih^-_{r,s}(q)&=&\frac{1}{(pp')^2\eta(q)}\bigg(\theta''_{pp'-sp-rp'}(q)
     -\theta''_{pp'+sp-rp'}(q)+(sp+rp')\theta'_{pp'-sp-rp'}(q)\nonumber\\
  &&\mbox{}\qquad\qquad\qquad
     +(sp-rp')\theta'_{pp'+sp-rp'}(q)+\frac{(sp+rp')^2-(pp')^2}{4}\,\theta_{pp'-sp-rp'}(q)
   \nonumber\\
 &&\mbox{}\qquad\qquad\qquad\qquad-\frac{(sp-rp')^2-(pp')^2}{4}\,\theta_{pp'+sp-rp'}(q)\bigg)
\eea
where the Dedekind eta function is defined in (\ref{eta}).
Uniqueness of the theta forms (\ref{tfab}) is obtained by imposing $ap'+bp\le pp'$.
Besides the identification of the ${\cal W}$-irreducible characters $\chih_{a,b}(q)$ in
(\ref{chihab}) with the theta forms of the same names in (\ref{tfab}),
the precise relations between the theta forms and our ${\cal W}$-irreducible characters are
\be
 \chih_{r,s}^+(q)\ =\ \chih_{r,2p'-s}(q)\ =\ \chih_{2p-r,s}(q),\qquad\quad
 \chih_{r,s}^-(q)\ =\ \chih_{r,3p'-s}(q)\ =\ \chih_{3p-r,s}(q)
\ee

\subsection{Embedding patterns}

We conjecture that every ${\cal W}$-indecomposable rank-2 representation
has an embedding pattern of one of the types 
\be
 \mbox{
 \begin{picture}(100,120)(0,0)
    \unitlength=1cm
  \thinlines
\put(-2.3,2){$\mathcal{E}(\D_h,\D_v):$}
\put(2,3.6){$\ketw{\D_v}$}
\put(0,2){$\ketw{\D_h}$}
\put(3.9,2){$\ketw{\D_h}$}
\put(2,0.5){$\ketw{\D_v}$}
\put(3.5,2.1){\vector(-1,0){2.1}}
\put(1.9,3.4){\vector(-4,-3){1.2}}
\put(4.1,1.7){\vector(-4,-3){1.2}}
\put(1.9,0.8){\vector(-4,3){1.2}}
\put(4.1,2.5){\vector(-4,3){1.2}}
 \end{picture}
}
\hspace{5.3cm}
 \mbox{
 \begin{picture}(100,120)(-10,0)
    \unitlength=1cm
  \thinlines
\put(-3,2){$\mathcal{E}(\D_h,\D_v;\D_c):$}
\put(2,3.6){$\ketw{\D_v}$}
\put(0,2){$\ketw{\D_h}$}
\put(3.9,2){$\ketw{\D_h}$}
\put(2,0.5){$\ketw{\D_v}$}
\put(3.5,2.1){\vector(-1,0){2.1}}
\put(1.9,3.4){\vector(-4,-3){1.2}}
\put(4.1,1.7){\vector(-4,-3){1.2}}
\put(1.9,0.8){\vector(-4,3){1.2}}
\put(4.1,2.5){\vector(-4,3){1.2}}
\put(2,1.35){$\ketw{\D_c}$}
\put(3.7,1.9){\vector(-2,-1){0.5}}
\put(1.8,1.6){\vector(-2,1){0.5}}
 \end{picture}
}
\label{E}
\ee 
where the horizontal arrows indicate the non-diagonal action of the Virasoro mode $L_0$.
Specifically, we conjecture that the ${\cal W}$-indecomposable rank-2 representations 
(\ref{r2}) enjoy the embedding patterns 
\bea
  &&\ketw{\R_{p,b}^{a,0}}\ \sim\ \Ec(\D_{p+a,b},\D_{3p-a,b};\D_{p-a,b}),\qquad\quad\ \
    \ketw{\R_{a,p'}^{0,b}}\ \sim\ \Ec(\D_{a,p'+b},\D_{a,3p'-b};\D_{a,p'-b})\nn
  &&\ketw{\R_{p,p'}^{a,0}}\ \sim\ \Ec(\D_{p+a,p'},\D_{3p-a,p'}),\qquad\qquad\qquad
    \ketw{\R_{p,p'}^{0,b}}\ \sim\ \Ec(\D_{p,p'+b},\D_{p,3p'-b})\nn
  &&\ketw{\R_{2p,s}^{a,0}}\ \sim\ \Ec(\D_{2p+a,s},\D_{2p-a,s}),\qquad\qquad\qquad
    \ketw{\R_{r,2p'}^{0,b}}\ \sim\ \Ec(\D_{r,2p'+b},\D_{r,2p'-b})
\eea
These embedding patterns demonstrate the inequivalence of
the various rank-2 representations despite the character identities (\ref{r2chid}).

We also conjecture that the
${\cal W}$-indecomposable rank-3 representations (\ref{r3}) have embedding structures
described by the patterns in (\ref{E}). Specifically, we conjecture that
\be
 \ketw{\R_{\kappa p,p'}^{a,b}}
  \ \sim\ \Ec\Big(\ketw{\R_{\kappa p,p'-b}^{a,0}},\ketw{\R_{(3-\kappa)p,b}^{a,0}}\Big)
  \ \sim\ \Ec\Big(\ketw{\R_{p-a,\kappa p'}^{0,b}},\ketw{\R_{a,(3-\kappa)p'}^{0,b}}\Big)
\label{Rr3E}
\ee
where the ${\cal W}$-irreducible representations $\ketw{\D_h}$ and $\ketw{\D_v}$
have been replaced by ${\cal W}$-indecomposable rank-2 representations.
It is noted that each of the $2(p-1)(p'-1)$ rank-3 representations is thus proposed to
be viewable in {\em two} different ways. This corresponds to viewing it as
an indecomposable `vertical' combination of `horizontal' rank-2 representations
$\ketw{\R^{a,0}}$ or as an indecomposable `horizontal' combination
of `vertical' rank-2 representations $\ketw{\R^{0,b}}$.
As in the case of rank-2 representations, the conjectured embedding
patterns (\ref{Rr3E}) demonstrate inequivalence of
the various rank-3 representations despite the character identities (\ref{r3chid}).

\subsection{Jordan-cell structures and refined characters}

To encode the Jordan-cell structure of a rank-$n$ representation of character $\chit(q)$, 
we introduce its `refined' character
\be
 \chit_{\mathrm{ref}}(q)\ =\ \sum_{j=1}^n\mathcal{J}_j\chit_{\mathrm{ref}}^j(q),\qquad\quad
 \chit(q)\ =\ \mathrm{Tr}\chit_{\mathrm{ref}}(q)\ =\ \sum_{j=1}^nj\chit_{\mathrm{ref}}^j(q)
\ee
where the $(j\times j)$-dimensional canonical Jordan cell $\mathcal{J}_j$ is defined by
\be
 \mathcal{J}_j\ =\ \begin{pmatrix} 1&1&&&& \\ 
                                                           &1&1&&& \\
                                                           &&\ddots&\ddots&& \\ 
                                                           &&&\ddots&\ddots&\\
                                                           &&&&1&1\\
                                                           &&&&&1\end{pmatrix},\qquad\quad
                           {(\mathcal{J}_j)_i}^{i'}\ =\ \delta_{i',i}+\delta_{i',i+1}
\ee
and has trace $\mathrm{Tr}\mathcal{J}_j=j$. Now, in the rank-$n$ representation under consideration,
the number of rank-$j$ Jordan cells at a given 
level $\ell$ is simply given by the multiplicity of $q^\ell$ in $\chit_{\mathrm{ref}}^j(q)$.
The mere number of Jordan cells at a given level $\ell$ is 
therefore given by the multiplicity of $q^\ell$ in $\chit_{\mathrm{ref}}^{\mathrm{tot}}(q)$ where
\be
 \chit_{\mathrm{ref}}^{\mathrm{tot}}(q)\ =\ \sum_{j=1}^n\chit_{\mathrm{ref}}^j(q)
\ee
For simplicity, we will omit the trivial matrix notation for $j=1$ and set $\mathcal{J}_1=1$.

We can also use this refined character notation when considering a decomposition of
a character in terms of irreducible characters, for example.
By 
\be
 2\ch_{r,s}(q)+\mathcal{J}_2\big(\ch_{r',s'}(q)+\ch_{r'',s''}(q)\big)
\ee
we thus mean a sum of 6 irreducible Virasoro characters
where a Jordan cell of rank 2 is formed between every pair of matching states
in the 2 modules labelled by $r',s'$ and between every pair of matching states
in the 2 modules labelled by $r'',s''$ while no state in the modules
labelled by $r,s$ is part of a non-trivial Jordan cell.
The refined characters of the ${\cal W}$-indecomposable rank-2 representations then read
\bea
 \chit_{\mathrm{ref}}\big[\ketw{\R_{\kappa p,s}^{a,0}}\big](q)
   &=&\delta_{\kappa,1}\big(1-\delta_{s,p'}\big)\ch_{p-a,s}(q)
  +2\sum_{k\in\mathbb{N}}(2k+1-\kappa)\ch_{(2k+2-\kappa)p-a,s}(q)\nn
   &&\hspace{1.2cm}
      +\ \mathcal{J}_2\sum_{k\in\mathbb{N}}(2k-2+\kappa)\ch_{(2k-2+\kappa)p+a,s}(q)\nn
 \chit_{\mathrm{ref}}\big[\ketw{\R_{r,\kappa p'}^{0,b}}\big](q)
   &=&\delta_{\kappa,1}\big(1-\delta_{r,p}\big)\ch_{r,p'-b}(q)
  +2\sum_{k\in\mathbb{N}}(2k+1-\kappa)\ch_{r,(2k+2-\kappa)p'-b}(q)\nn
   &&\hspace{1.2cm}
      +\ \mathcal{J}_2\sum_{k\in\mathbb{N}}(2k-2+\kappa)\ch_{r,(2k-2+\kappa)p'+b}(q)
\eea
which can be re-expressed in terms of ${\cal W}$-irreducible characters as
\bea
 \chit_{\mathrm{ref}}\big[\ketw{\R_{\kappa p,s}^{a,0}}\big](q)
   &=& \delta_{\kappa,1}\big(1-\delta_{s,p'}\big)\chih_{p-a,s}(q)+2\chih_{(4-\kappa)p-a,s}(q)
    +\mathcal{J}_2\chih_{\kappa p+a,s}(q)\nn
 \chit_{\mathrm{ref}}\big[\ketw{\R_{r,\kappa p'}^{0,b}}\big](q)&=&
    \delta_{\kappa,1}\big(1-\delta_{r,p}\big)\chih_{r,p'-b}(q)+2\chih_{r,(4-\kappa)p'-b}(q)
    +\mathcal{J}_2\chih_{r,\kappa p'+b}(q)
\label{r2ref}
\eea
It follows that the refined character components are given by
\bea
 \chit_{\mathrm{ref}}^1\big[\ketw{\R_{\kappa p,s}^{a,0}}\big](q)
   &=&\delta_{\kappa,1}\big(1-\delta_{s,p'}\big)\chih_{p-a,s}(q)+2\chih_{(4-\kappa)p-a,s}(q)\nn
 \chit_{\mathrm{ref}}^2\big[\ketw{\R_{\kappa p,s}^{a,0}}\big](q)
   &=&\chih_{\kappa p+a,s}(q)
\eea
and
\bea
 \chit_{\mathrm{ref}}^1\big[\ketw{\R_{r,\kappa p'}^{0,b}}\big](q)
  &=&\delta_{\kappa,1}\big(1-\delta_{r,p}\big)\chih_{r,p'-b}(q)+2\chih_{r,(4-\kappa)p'-b}(q)\nn
 \chit_{\mathrm{ref}}^2\big[\ketw{\R_{r,\kappa p'}^{0,b}}\big](q)
  &=&\chih_{r,\kappa p'+b}(q)
\eea
We note that the refined character expressions contain enough information to
distinguish between the different rank-2 representations. That is, the distinctions
can be made by solely emphasizing the Jordan-cell structures without further reference to
the complete embedding patterns. We also note that the refined character components
are related to each other
\be
 \chit_{\mathrm{ref}}^1\big[\ketw{\R_{\kappa p,b}^{a,0}}\big](q)
   \ =\ \chit_{\mathrm{ref}}^1\big[\ketw{\R_{a,\kappa p'}^{0,b}}\big](q),\qquad\quad
 \chit_{\mathrm{ref}}^2\big[\ketw{\R_{\kappa p,b}^{a,0}}\big](q)
   \ =\ \chit_{\mathrm{ref}}^2\big[\ketw{\R_{p-a,\kappa p'}^{0,p'-b}}\big](q)
\ee
from which it follows, in particular, that the refined characters themselves satisfy the relations
\be
 \chit_{\mathrm{ref}}\big[\ketw{\R_{\kappa p,b}^{a,0}}\big](q)
  +\chit_{\mathrm{ref}}\big[\ketw{\R_{\kappa p,p'-b}^{p-a,0}}\big](q)
 \ =\ 
 \chit_{\mathrm{ref}}\big[\ketw{\R_{a,\kappa p'}^{0,b}}\big](q)
  +\chit_{\mathrm{ref}}\big[\ketw{\R_{p-a,\kappa p'}^{0,p'-b}}\big](q)
\ee
and hence
\be
 \sum_{a=1}^{p-1}\sum_{b=1}^{p'-1}\chit_{\mathrm{ref}}\big[\ketw{\R_{\kappa p,b}^{a,0}}\big](q)
  \ =\ 
 \sum_{a=1}^{p-1}\sum_{b=1}^{p'-1}\chit_{\mathrm{ref}}\big[\ketw{\R_{a,\kappa p'}^{0,b}}\big](q)
\ee
 
Refinements of the rank-3 characters similar to the refined characters 
(\ref{r2ref}) follow from the conjectured embedding
patterns (\ref{Rr3E}). Converting the two rank-2 Jordan cells linked by a horizontal arrow 
in (\ref{Rr3E}) into a rank-3 and a rank-1 Jordan cell \cite{EF06,RP0707,RP0804}, 
we arrive at the refined characters
\bea
 \chit_{\mathrm{ref}}\big[\ketw{\R_{\kappa p,p'}^{a,b}}\big](q)
   &=& \delta_{\kappa,1}\mathcal{J}_2\chih_{a,b}(q)+2\delta_{\kappa,2}\chih_{p-a,b}(q)
    +\big\{\mathcal{J}_3+1\big\}\chih_{\kappa p+a,p'-b}(q)\nn
   &+&4\chih_{\kappa p+p-a,b}(q)+2\mathcal{J}_2\chih_{(3-\kappa)p+a,b}(q)
    +2\mathcal{J}_2\chih_{a,(3-\kappa)p'+b}(q)\nn
   &=&\sum_{j=1}^3\mathcal{J}_j\chit_{\mathrm{ref}}^j\big[\ketw{\R_{\kappa p,p'}^{a,b}}\big](q)
\eea
where the refined character components are given by
\bea
 \chit_{\mathrm{ref}}^1\big[\ketw{\R_{\kappa p,p'}^{a,b}}\big](q)&=&2\delta_{\kappa,2}\chih_{p-a,b}(q)
   +\chih_{\kappa p+a,p'-b}(q)+4\chih_{\kappa p+p-a,b}(q)  \nn
 \chit_{\mathrm{ref}}^2\big[\ketw{\R_{\kappa p,p'}^{a,b}}\big](q)&=&\delta_{\kappa,1}\chih_{a,b}(q)
   +2\chih_{(3-\kappa)p+a,b}(q)+2\chih_{a,(3-\kappa)p'+b}(q)  \nn
 \chit_{\mathrm{ref}}^3\big[\ketw{\R_{\kappa p,p'}^{a,b}}\big](q)&=& \chih_{\kappa p+a,p'-b}(q)
\eea
In order to demonstrate inequivalence of the ${\cal W}$-indecomposable rank-3 representations,
it suffices to focus on the presence of rank-3 Jordan cells.
This follows from the fact that the ${\cal W}$-irreducible subfactors with characters
$\chih_{\kappa p+a,p'-b}(q)$ are distinct for every distinct choice of
$\kappa, a, b$ (of which there are $2(p-1)(p'-1)$ possibilities),

Due to the various character relations satisfied by the ${\cal W}$-indecomposable rank-2
representations, it follows from the embedding patterns (\ref{Rr3E}) of the ${\cal W}$-indecomposable
rank-3 representations that there are many relations for the (refined) 
rank-3 characters as well. As they do not seem to shed new light on the structure of the
rank-3 representations, they will not be discussed any further here.

\subsection{${\cal W}$-extended fusion algebra}
\label{SectionFusAlg}

We denote the fusion product in the ${\cal W}$-extended picture by $\fus$ and
reserve the symbol $\otimes$ for the fusion product in the Virasoro picture.
The fusion rules underlying the fusion algebra in the ${\cal W}$-extended picture ${\cal WLM}(p,p')$ 
\bea
 \big\langle\ketw{\kappa p,s},\ketw{r,\kappa' p'},\ketw{\R_{\kappa p,s}^{a,0}},
  \ketw{\R_{r,\kappa' p'}^{0,b}},\ketw{\R_{\kappa p,\kappa' p'}^{a,b}}  \big\rangle_{p,p'}
\label{Wfus}
\eea
are summarized in the following. Here it is recalled that the
various ${\cal W}$-indecomposable representations are subject to (\ref{identi}).
The fusion of two ${\cal W}$-indecomposable rank-1 representations is given by
\bea
 \ketw{\kappa p,s}\fus\ketw{\kappa'p,s'}&=&\bigoplus_{\al}^{p-1}\Big\{
  \!\bigoplus_{j=|s-s'|+1,\ \!{\rm by}\ \!2}^{p'-|p'-s-s'|-1}
  \!\!\!\ketw{\R_{(\kappa\cdot\kappa')p,j}^{\al,0}}
    \oplus\!\bigoplus_{\beta}^{s+s'-p'-1}
    \!\ketw{\R_{\kappa p,\kappa'p'}^{\al,\beta}}
    \Big\}   \nn
 \ketw{\kappa p,s}\fus\ketw{r,\kappa'p'}&=&\bigoplus_{\al}^{r-1}
  \Big\{\bigoplus_{\beta}^{s-1}
     \ketw{\R_{\kappa p,\kappa'p'}^{\al,\beta}}\Big\}\nn
 \ketw{r,\kappa p'}\fus\ketw{r',\kappa'p'}&=&\bigoplus_{\beta}^{p'-1}\Big\{
   \!\bigoplus_{j=|r-r'|+1,\ \!{\rm by}\ \!2}^{p-|p-r-r'|-1}
  \!\!\!\ketw{\R_{j,(\kappa\cdot\kappa')p'}^{0,\beta}}
  \oplus\!\bigoplus_{\al}^{r+r'-p-1}
    \!\ketw{\R_{\kappa p,\kappa'p'}^{\al,\beta}}
    \Big\}
\label{fus11}
\eea
The fusion of a ${\cal W}$-indecomposable rank-1 representation with a 
${\cal W}$-indecomposable rank-2 representation is given by
\bea
 \ketw{\kappa p,s}\fus\ketw{\R_{\kappa'p,s'}^{a,0}}
   &=&\!\bigoplus_{j=|s-s'|+1,\ \!{\rm by}\ \!2}^{p'-|p'-s-s'|-1}
  \!\!\Big\{\bigoplus_{\al}^{p-a-1}\!\!2\ketw{\R_{(\kappa\cdot\kappa')p,j}^{\al,0}}\oplus
    \bigoplus_{\al}^{a-1}2\ketw{\R_{(2\cdot\kappa\cdot\kappa')p,j}^{\al,0}}\Big\}  \nn
  &\oplus&\bigoplus_{\beta}^{s+s'-p'-1}
  \!\!\Big\{\bigoplus_{\al}^{p-a-1}\!\!2\ketw{\R_{\kappa p,\kappa'p'}^{\al,\beta}}\oplus
    \bigoplus_{\al}^{a-1}2\ketw{\R_{\kappa p,(2\cdot\kappa')p'}^{\al,\beta}}\Big\}  \nn
 \ketw{\kappa p,s}\fus\ketw{\R_{r,\kappa'p'}^{0,b}}
  &=&\bigoplus_{\al}^{r-1}\Big\{
    \!\bigoplus_{\beta=|b-s|+1,\ \!{\rm by}\ \!2}^{p'-|p'-s-b|-1}
      \!\!\!\!\ketw{\R_{\kappa p,\kappa'p'}^{\al,\beta}}
     \oplus\!\bigoplus_{\beta}^{s-b-1}\!2\ketw{\R_{\kappa p,\kappa'p'}^{\al,\beta}}\oplus
     \!\!\!\bigoplus_{\beta}^{b+s-p'-1}
        \!\!\!2\ketw{\R_{\kappa p,(2\cdot\kappa')p'}^{\al,\beta}}\Big\}   \nn
 \ketw{r,\kappa p'}\fus\ketw{\R_{\kappa'p,s}^{a,0}}
  &=&\bigoplus_{\beta}^{s-1}\Big\{
    \!\bigoplus_{\al=|a-r|+1,\ \!{\rm by}\ \!2}^{p-|p-r-a|-1}
      \!\!\!\!\ketw{\R_{\kappa p,\kappa'p'}^{\al,\beta}}
    \oplus\!\bigoplus_{\al}^{r-a-1}\!2\ketw{\R_{\kappa p,\kappa'p'}^{\al,\beta}}\oplus 
     \!\!\!\bigoplus_{\al}^{a+r-p-1}
        \!\!\!2\ketw{\R_{\kappa p,(2\cdot\kappa')p'}^{\al,\beta}}\Big\}   \nn
 \!\ketw{r,\kappa p'}\fus\ketw{\R_{r',\kappa'p'}^{0,b}}
  &=&\!\bigoplus_{j=|r-r'|+1,\ \!{\rm by}\ \!2}^{p-|p-r-r'|-1}
  \!\!\Big\{\bigoplus_{\beta}^{p'-b-1}\!\!2\ketw{\R_{j,(\kappa\cdot\kappa')p'}^{0,\beta}}\oplus
    \bigoplus_{\beta}^{b-1}2\ketw{\R_{j,(2\cdot\kappa\cdot\kappa')p'}^{0,\beta}}\Big\}  \nn
  &\oplus&\bigoplus_{\al}^{r+r'-p-1}
  \!\!\Big\{\bigoplus_{\beta}^{p'-b-1}\!\!2\ketw{\R_{\kappa p,\kappa'p'}^{\al,\beta}}\oplus
    \bigoplus_{\beta}^{b-1}2\ketw{\R_{\kappa p,(2\cdot\kappa')p'}^{\al,\beta}}\Big\}   
\label{fus12}
\eea
The fusion of a ${\cal W}$-indecomposable rank-1 representation with a 
${\cal W}$-indecomposable rank-3 representation is given by
\bea
 \ketw{\kappa p,s}\fus\ketw{\R_{p,\kappa'p'}^{a,b}}
  &=&\!\!\!\bigoplus_{\al}^{p-a-1}\!\!\Big\{
     \!\bigoplus_{\beta=|b-s|+1,\ \!{\rm by}\ \!2}^{p'-|p'-s-b|-1}
      \!\!\!\!2\ketw{\R_{\kappa p,\kappa'p'}^{\al,\beta}}
   \oplus\!\bigoplus_{\beta}^{s-b-1}\!\!4\ketw{\R_{\kappa p,\kappa'p'}^{\al,\beta}}
     \oplus\!\!\!\!\bigoplus_{\beta}^{b+s-p'-1}
        \!\!\!\!4\ketw{\R_{\kappa p,(2\cdot\kappa')p'}^{\al,\beta}}\Big\}\nn
 &\oplus&\!\!\!\bigoplus_{\al}^{a-1}\!\Big\{
     \!\!\bigoplus_{\beta=|b-s|+1,\ \!{\rm by}\ \!2}^{p'-|p'-s-b|-1}
      \!\!\!\!2\ketw{\R_{\kappa p,(2\cdot\kappa')p'}^{\al,\beta}}
   \oplus\!\bigoplus_{\beta}^{s-b-1}\!\!4\ketw{\R_{\kappa p,(2\cdot\kappa')p'}^{\al,\beta}}
     \oplus\!\!\!\!\!\bigoplus_{\beta}^{b+s-p'-1}
        \!\!\!\!\!4\ketw{\R_{\kappa p,\kappa'p'}^{\al,\beta}}\Big\}  \nn
 \ketw{r,\kappa p'}\fus\ketw{\R_{p,\kappa'p'}^{a,b}}
  &=&\!\!\!\bigoplus_{\beta}^{p'-b-1}\!\!\!\Big\{
     \!\bigoplus_{\al=|a-r|+1,\ \!{\rm by}\ \!2}^{p-|p-r-a|-1}
      \!\!\!\!2\ketw{\R_{\kappa p,\kappa'p'}^{\al,\beta}}
  \oplus\!\!\bigoplus_{\al}^{r-a-1}\!\!4\ketw{\R_{\kappa p,\kappa'p'}^{\al,\beta}}
     \oplus\!\!\!\!\bigoplus_{\al}^{a+r-p-1}
        \!\!\!\!4\ketw{\R_{\kappa p,(2\cdot\kappa')p'}^{\al,\beta}}\Big\}\nn
 &\oplus&\!\!\!\bigoplus_{\beta}^{b-1}\!\Big\{
     \!\bigoplus_{\al=|a-r|+1,\ \!{\rm by}\ \!2}^{p-|p-r-a|-1}
      \!\!\!\!2\ketw{\R_{\kappa p,(2\cdot\kappa')p'}^{\al,\beta}}
  \oplus\!\!\bigoplus_{\al}^{r-a-1}\!\!\!4\ketw{\R_{\kappa p,(2\cdot\kappa')p'}^{\al,\beta}}
     \oplus\!\!\!\!\!\bigoplus_{\al}^{a+r-p-1}
        \!\!\!\!\!4\ketw{\R_{\kappa p,\kappa'p'}^{\al,\beta}}\Big\}
     \nn
\label{fus13}
\eea
The fusion of two ${\cal W}$-indecomposable rank-2 representations is given by
\bea
 \ketw{\R_{\kappa p,s}^{a,0}}\fus\ketw{\R_{\kappa'p,s'}^{a',0}}
  &=&\!\bigoplus_{j=|s-s'|+1,\ \!{\rm by}\ \!2}^{p'-|p'-s-s'|-1}
   \!\!\Big\{\bigoplus_{\al}^{p-|a-a'|-1}
     \!\!2\ketw{\R_{(\kappa\cdot\kappa')p,j}^{\al,0}}
     \oplus\!\bigoplus_{\al}^{|p-a-a'|-1}
     \!\!2\ketw{\R_{(\kappa\cdot\kappa')p,j}^{\al,0}}\nn
   &&\qquad\oplus\bigoplus_{\al}^{p-|p-a-a'|-1}
     \!\!\!2\ketw{\R_{(2\cdot\kappa\cdot\kappa')p,j}^{\al,0}}
     \oplus\bigoplus_{\al}^{|a-a'|-1}
     \!2\ketw{\R_{(2\cdot\kappa\cdot\kappa')p,j}^{\al,0}}\Big\}\nn
&\oplus&\bigoplus_{\beta}^{s+s'-p'-1}   \!\!\Big\{\bigoplus_{\al}^{p-|a-a'|-1}
     \!\!2\ketw{\R_{\kappa p,\kappa'p'}^{\al,\beta}}
     \oplus\bigoplus_{\al}^{|p-a-a'|-1}
     \!\!2\ketw{\R_{\kappa p,\kappa'p'}^{\al,\beta}}\nn
   &&\qquad\oplus\bigoplus_{\al}^{p-|p-a-a'|-1}
     \!\!\!2\ketw{\R_{\kappa p,(2\cdot\kappa')p'}^{\al,\beta}}
     \oplus\bigoplus_{\al}^{|a-a'|-1}
     \!2\ketw{\R_{\kappa p,(2\cdot\kappa')p'}^{\al,\beta}}\Big\}\nn
 \ketw{\R_{\kappa p,s}^{a,0}}\fus\ketw{\R_{r,\kappa'p'}^{0,b}}
  &=&
   \!\!\bigoplus_{\al=|a-r|+1,\ \!{\rm by}\ \!2}^{p-|p-r-a|-1}
   \!\Big\{\bigoplus_{\beta=|b-s|+1,\ \!{\rm by}\ \!2}^{p'-|p'-s-b|-1}
    \!\!\!\!\ketw{\R_{\kappa p,\kappa'p'}^{\al,\beta}}\Big\}
    \nn
 &\oplus&  
     \!\!\bigoplus_{\al=|a-r|+1,\ \!{\rm by}\ \!2}^{p-|p-r-a|-1}
   \!\Big\{  \bigoplus_{\beta}^{s-b-1}2\ketw{\R_{\kappa p,\kappa'p'}^{\al,\beta}}\Big\}
    \oplus\bigoplus_{\beta=|b-s|+1,\ \!{\rm by}\ \!2}^{p'-|p'-s-b|-1}
    \!\Big\{  \bigoplus_{\al}^{r-a-1}2\ketw{\R_{\kappa p,\kappa'p'}^{\al,\beta}}\Big\}
    \nn
  &\oplus&\!\bigoplus_{\al}^{r-a-1}\!\Big\{ 
     \bigoplus_{\beta}^{s-b-1}4\ketw{\R_{\kappa p,\kappa'p'}^{\al,\beta}}\Big\}
      \oplus\!\bigoplus_{\al}^{a+r-p-1}\!\Big\{
     \bigoplus_{\beta}^{b+s-p'-1}\!4\ketw{\R_{\kappa p,\kappa'p'}^{\al,\beta}}\Big\}
     \nn
 &\oplus&\!\bigoplus_{\al}^{a+r-p-1}\!\Big\{
   \bigoplus_{\beta=|b-s|+1,\ \!{\rm by}\ \!2}^{p'-|p'-s-b|-1}
     \!\!2\ketw{\R_{\kappa p,(2\cdot\kappa')p'}^{\al,\beta}}\oplus
      \bigoplus_{\beta}^{s-b-1}\!4\ketw{\R_{\kappa p,(2\cdot\kappa')p'}^{\al,\beta}}\Big\}
      \nn
 &\oplus&\!\bigoplus_{\beta}^{b+s-p'-1}\!\Big\{
  \bigoplus_{\al=|a-r|+1,\ \!{\rm by}\ \!2}^{p-|p-r-a|-1}
    \!\!2\ketw{\R_{\kappa p,(2\cdot\kappa')p'}^{\al,\beta}}\oplus
     \bigoplus_{\beta}^{r-a-1}\!4\ketw{\R_{\kappa p,(2\cdot\kappa')p'}^{\al,\beta}}\Big\}
     \nn
     \nn
 \ketw{\R_{r,\kappa p'}^{0,b}}\fus\ketw{\R_{r',\kappa'p'}^{0,b'}}
  &=&\!\!\bigoplus_{j=|r-r'|+1,\ \!{\rm by}\ \!2}^{p-|p-r-r'|-1}
   \!\!\Big\{\bigoplus_{\beta}^{p'-|b-b'|-1}
     \!\!2\ketw{\R_{j,(\kappa\cdot\kappa')p'}^{0,\beta}}
     \oplus\bigoplus_{\beta}^{|p'-b-b'|-1}
     \!\!2\ketw{\R_{j,(\kappa\cdot\kappa')p'}^{0,\beta}}\nn
   &&\qquad\oplus\bigoplus_{\beta}^{p'-|p'-b-b'|-1}
     \!\!\!2\ketw{\R_{j,(2\cdot\kappa\cdot\kappa')p'}^{0,\beta}}
     \oplus\bigoplus_{\beta}^{|b-b'|-1}
     \!2\ketw{\R_{j,(2\cdot\kappa\cdot\kappa')p'}^{0,\beta}}\Big\}\nn
&\oplus&\bigoplus_{\al}^{r+r'-p-1}   \!\Big\{\bigoplus_{\beta}^{p'-|b-b'|-1}
     \!\!2\ketw{\R_{\kappa p,\kappa'p'}^{\al,\beta}}
     \oplus\!\bigoplus_{\beta}^{|p'-b-b'|-1}
     \!\!2\ketw{\R_{\kappa p,\kappa'p'}^{\al,\beta}}\nn
   &&\qquad\oplus\!\bigoplus_{\beta}^{p'-|p'-b-b'|-1}
     \!\!\!2\ketw{\R_{\kappa p,(2\cdot\kappa')p'}^{\al,\beta}}
     \oplus\bigoplus_{\beta}^{|b-b'|-1}
     \!2\ketw{\R_{\kappa p,(2\cdot\kappa')p'}^{\al,\beta}}\Big\}  
\label{fus22}
\eea
The fusion of a ${\cal W}$-indecomposable rank-2 representation with a 
${\cal W}$-indecomposable rank-3 representation is given by
\bea
 \ketw{\R_{\kappa p,s}^{a,0}}\fus\ketw{\R_{p,\kappa'p'}^{a',b'}}
  &=&\!\!\bigoplus_{\beta=|b'-s|+1,\ \!{\rm by}\ \!2}^{p'-|p'-s-b'|-1}\!\!\Big\{
   \bigoplus_{\al}^{p-|a-a'|-1}\!\!\!2\ketw{\R_{\kappa p,\kappa'p'}^{\al,\beta}}
   \oplus\!\bigoplus_{\al}^{|p-a-a'|-1}\!\!\!2\ketw{\R_{\kappa p,\kappa'p'}^{\al,\beta}}\Big\}
  \nn
  &\oplus&\bigoplus_{\beta}^{s-b'-1}\!\Big\{
   \bigoplus_{\al}^{p-|a-a'|-1}\!\!\!4\ketw{\R_{\kappa p,\kappa'p'}^{\al,\beta}}
   \oplus\!\bigoplus_{\al}^{|p-a-a'|-1}\!\!\!4\ketw{\R_{\kappa p,\kappa'p'}^{\al,\beta}}\Big\}
  \nn
  &\oplus&\bigoplus_{\beta}^{b'+s-p'-1}\!\Big\{
   \bigoplus_{\al}^{p-|p-a-a'|-1}\!\!\!4\ketw{\R_{\kappa p,\kappa'p'}^{\al,\beta}}\oplus
   \bigoplus_{\al}^{|a-a'|-1}\!4\ketw{\R_{\kappa p,\kappa'p'}^{\al,\beta}}\Big\}
 \nn
 &\oplus&\!\!\bigoplus_{\beta=|b'-s|+1,\ \!{\rm by}\ \!2}^{p'-|p'-s-b'|-1}\!\!\Big\{
   \bigoplus_{\al}^{p-|p-a-a'|-1}\!\!\!2\ketw{\R_{\kappa p,(2\cdot\kappa')p'}^{\al,\beta}}
   \oplus\bigoplus_{\al}^{|a-a'|-1}\!2\ketw{\R_{\kappa p,(2\cdot\kappa')p'}^{\al,\beta}}\Big\}
 \nn
 &\oplus&\bigoplus_{\beta}^{s-b'-1}\!\Big\{
   \bigoplus_{\al}^{p-|p-a-a'|-1}\!\!\!4\ketw{\R_{\kappa p,(2\cdot\kappa')p'}^{\al,\beta}}\oplus
   \bigoplus_{\al}^{|a-a'|-1}\!4\ketw{\R_{\kappa p,(2\cdot\kappa')p'}^{\al,\beta}}\Big\}
 \nn
 &\oplus&\bigoplus_{\beta}^{b'+s-p'-1}\!\Big\{
   \bigoplus_{\al}^{p-|a-a'|-1}\!\!\!4\ketw{\R_{\kappa p,(2\cdot\kappa')p'}^{\al,\beta}}
   \oplus\!\bigoplus_{\al}^{|p-a-a'|-1}\!\!\!4\ketw{\R_{\kappa p,(2\cdot\kappa')p'}^{\al,\beta}}\Big\} 
 \nn 
 \ketw{\R_{r,\kappa p'}^{0,b}}\fus\ketw{\R_{p,\kappa'p'}^{a',b'}}
  &=&\!\!\bigoplus_{\al=|a'-r|+1,\ \!{\rm by}\ \!2}^{p-|p-r-a'|-1}\!\!\Big\{
   \bigoplus_{\beta}^{p'-|b-b'|-1}\!\!\!2\ketw{\R_{\kappa p,\kappa'p'}^{\al,\beta}}
   \oplus\!\bigoplus_{\beta}^{|p'-b-b'|-1}\!\!\!2\ketw{\R_{\kappa p,\kappa'p'}^{\al,\beta}}\Big\}
  \nn
  &\oplus&\bigoplus_{\al}^{r-a'-1}\!\Big\{
   \bigoplus_{\beta}^{p'-|b-b'|-1}\!\!\!4\ketw{\R_{\kappa p,\kappa'p'}^{\al,\beta}}
   \oplus\!\bigoplus_{\beta}^{|p'-b-b'|-1}\!\!\!4\ketw{\R_{\kappa p,\kappa'p'}^{\al,\beta}}\Big\}
  \nn
  &\oplus&\bigoplus_{\al}^{a'+r-p-1}\!\Big\{
   \bigoplus_{\beta}^{p'-|p'-b-b'|-1}\!\!\!4\ketw{\R_{\kappa p,\kappa'p'}^{\al,\beta}}\oplus
   \bigoplus_{\beta}^{|b-b'|-1}\!4\ketw{\R_{\kappa p,\kappa'p'}^{\al,\beta}}\Big\}
 \nn
 &\oplus&\!\!\bigoplus_{\al=|a'-r|+1,\ \!{\rm by}\ \!2}^{p-|p-r-a'|-1}\!\!\Big\{
   \bigoplus_{\beta}^{p'-|p'-b-b'|-1}\!\!\!2\ketw{\R_{\kappa p,(2\cdot\kappa')p'}^{\al,\beta}}
   \oplus\bigoplus_{\beta}^{|b-b'|-1}\!2\ketw{\R_{\kappa p,(2\cdot\kappa')p'}^{\al,\beta}}\Big\}
 \nn
 &\oplus&\bigoplus_{\al}^{r-a'-1}\!\Big\{
   \bigoplus_{\beta}^{p'-|p'-b-b'|-1}\!\!\!4\ketw{\R_{\kappa p,(2\cdot\kappa')p'}^{\al,\beta}}\oplus
   \bigoplus_{\beta}^{|b-b'|-1}\!4\ketw{\R_{\kappa p,(2\cdot\kappa')p'}^{\al,\beta}}\Big\}
 \nn
 &\oplus&\bigoplus_{\al}^{a'+r-p-1}\!\Big\{
   \bigoplus_{\beta}^{p'-|b-b'|-1}\!\!\!4\ketw{\R_{\kappa p,(2\cdot\kappa')p'}^{\al,\beta}}
   \oplus\!\bigoplus_{\beta}^{|p'-b-b'|-1}\!\!\!4\ketw{\R_{\kappa p,(2\cdot\kappa')p'}^{\al,\beta}}\Big\}  
\label{fus23}
\eea
Finally, the fusion of two ${\cal W}$-indecomposable rank-3 representations is given by
\bea
 \ketw{\R_{\kappa p,p'}^{a,b}}\fus\ketw{\R_{p,\kappa'p'}^{a',b'}}
 &=&\!\bigoplus_{\al}^{p-|a-a'|-1}\!\!\Big\{
  \bigoplus_{\beta}^{p'-|b-b'|-1}\!\!\!4\ketw{\R_{\kappa p,\kappa'p'}^{\al,\beta}}\Big\}
   \oplus\!\!\bigoplus_{\al}^{|p-a-a'|-1}\!\!\Big\{
  \bigoplus_{\beta}^{|p'-b-b'|-1}\!\!\!4\ketw{\R_{\kappa p,\kappa'p'}^{\al,\beta}}\Big\}
 \nn
 &\oplus&\!\bigoplus_{\al}^{p-|a-a'|-1}\!\!\Big\{
  \bigoplus_{\beta}^{|p'-b-b'|-1}\!\!\!4\ketw{\R_{\kappa p,\kappa'p'}^{\al,\beta}}\Big\}
   \oplus\!\!\bigoplus_{\al}^{|p-a-a'|-1}\!\!\Big\{
  \bigoplus_{\beta}^{p'-|b-b'|-1}\!\!\!4\ketw{\R_{\kappa p,\kappa'p'}^{\al,\beta}}\Big\}
 \nn
 &\oplus&\!\bigoplus_{\al}^{p-|p-a-a'|-1}\!\!\Big\{
  \bigoplus_{\beta}^{p'-|p'-b-b'|-1}\!\!\!4\ketw{\R_{\kappa p,\kappa'p'}^{\al,\beta}}\Big\}
   \oplus\!\!\bigoplus_{\al}^{|a-a'|-1}\!\!\Big\{
  \bigoplus_{\beta}^{|b-b'|-1}\!\!\!4\ketw{\R_{\kappa p,\kappa'p'}^{\al,\beta}}\Big\}
 \nn
 &\oplus&\!\bigoplus_{\al}^{p-|p-a-a'|-1}\!\!\Big\{
  \bigoplus_{\beta}^{|b-b'|-1}\!\!\!4\ketw{\R_{\kappa p,\kappa'p'}^{\al,\beta}}\Big\}
   \oplus\!\!\bigoplus_{\al}^{|a-a'|-1}\!\!\Big\{
  \bigoplus_{\beta}^{p'-|p'-b-b'|-1}\!\!\!4\ketw{\R_{\kappa p,\kappa'p'}^{\al,\beta}}\Big\}
 \nn
 &\oplus&\!\bigoplus_{\al}^{p-|a-a'|-1}\!\!\Big\{
  \bigoplus_{\beta}^{p'-|p'-b-b'|-1}\!\!\!4\ketw{\R_{\kappa p,(2\cdot\kappa')p'}^{\al,\beta}}
  \oplus\!\!\bigoplus_{\beta}^{|b-b'|-1}\!\!\!4\ketw{\R_{\kappa p,(2\cdot\kappa')p'}^{\al,\beta}}\Big\}
 \nn
 &\oplus&\!\bigoplus_{\al}^{|p-a-a'|-1}\!\!\Big\{
    \bigoplus_{\beta}^{p'-|p'-b-b'|-1}\!\!\!4\ketw{\R_{\kappa p,(2\cdot\kappa')p'}^{\al,\beta}}
  \oplus\!\!\bigoplus_{\beta}^{|b-b'|-1}\!\!\!4\ketw{\R_{\kappa p,(2\cdot\kappa')p'}^{\al,\beta}}\Big\}
 \nn
 &\oplus&\!\bigoplus_{\beta}^{p'-|b-b'|-1}\!\!\Big\{
  \bigoplus_{\al}^{p-|p-a-a'|-1}\!\!\!4\ketw{\R_{\kappa p,(2\cdot\kappa')p'}^{\al,\beta}}
  \oplus\!\!\bigoplus_{\al}^{|a-a'|-1}\!\!\!4\ketw{\R_{\kappa p,(2\cdot\kappa')p'}^{\al,\beta}}\Big\}
 \nn
 &\oplus&\!\bigoplus_{\beta}^{|p'-b-b'|-1}\!\!\Big\{
    \bigoplus_{\al}^{p-|p-a-a'|-1}\!\!\!4\ketw{\R_{\kappa p,(2\cdot\kappa')p'}^{\al,\beta}}
  \oplus\!\!\bigoplus_{\al}^{|a-a'|-1}\!\!\!4\ketw{\R_{\kappa p,(2\cdot\kappa')p'}^{\al,\beta}}\Big\}
\label{fus33}
\eea
This fusion algebra is both associative and commutative, while there is no identity for $p>1$.
For $p=1$, the ${\cal W}$-irreducible representation $\ketw{1,1}$ is the identity.

\subsection{${\cal W}$-projective representations and their fusion algebra}
\label{SectionProj}

Here it suffices to characterize a ${\cal W}$-projective representation as a ${\cal W}$-indecomposable
representation which does not appear as a subfactor of any ${\cal W}$-indecomposable
representation different from itself. It follows that   
there are 2 ${\cal W}$-projective representations of rank 1
\be
 \ketw{\kappa p,p'}\ \equiv\ \ketw{p,\kappa p'}
\label{proj1}
\ee
$2p+2p'-4$ ${\cal W}$-projective representations of rank 2
\be
 \ketw{\R_{\kappa p,p'}^{a,0}},\qquad\ketw{\R_{p,\kappa p'}^{0,b}}
\label{proj2}
\ee
and $2(p-1)(p'-1)$ ${\cal W}$-projective representations of rank 3
\be
 \ketw{\R_{\kappa p,p'}^{a,b}}\ \equiv\ \ketw{\R_{p,\kappa p'}^{a,b}}
\label{proj3}
\ee
This gives a total of $N_{\mathrm{proj}}(p,p')$ ${\cal W}$-projective representations
where
\be
 N_{\mathrm{proj}}(p,p')\ =\ 2pp'
\label{Nproj}
\ee
It also follows that every ${\cal W}$-indecomposable rank-3 representation is a ${\cal W}$-projective
representation, while the 2 ${\cal W}$-indecomposable rank-1 representations
(\ref{proj1}) are both ${\cal W}$-irreducible and ${\cal W}$-projective.

We will refer to a character of a ${\cal W}$-projective representation as a 
${\cal W}$-projective character.
Due to the character identities (\ref{r2chid}), the number of linearly independent 
${\cal W}$-projective characters is smaller than $N_{\mathrm{proj}}(p,p')$ and is given by
\be
 \frac{1}{2}(p+1)(p'+1)
\label{projch}
\ee
This agrees with the counting of ${\cal W}$-projective characters in \cite{FGST06b}.
It is noted that the fermionic character expressions appearing in (\ref{r2q}) and (\ref{r3q})
exactly correspond to the ${\cal W}$-projective representations of rank 2 in
(\ref{proj2}) and rank 3 in (\ref{proj3}), respectively.

We find that the ${\cal W}$-indecomposable rank-1 representations
\be
 \ketw{\kappa p,b},\qquad \ketw{a,\kappa p'}
\ee
appear as subfactors of the ${\cal W}$-projective rank-2 representations (\ref{proj2}),
while the ${\cal W}$-indecomposable rank-2 representations
\be
 \ketw{\R_{\kappa p,b}^{a,0}},\qquad \ketw{\R_{a,\kappa p'}^{0,b}}
\ee
appear as subfactors of the ${\cal W}$-projective rank-3 representations (\ref{proj3}).
This exhausts the set of ${\cal W}$-indecomposable representations appearing
in the fusion algebra (\ref{Wfus}). The additional ${\cal W}$-irreducible representations
introduced in Section \ref{SectionIrrSub}
\be
 \ketw{a,b},\qquad \ketw{2p-a,b}\ \equiv\ \ketw{a,2p'-b},\qquad\ketw{3p-a,b}\ \equiv\ \ketw{a,3p'-b}
\ee
all appear as subfactors of the ${\cal W}$-projective rank-3 representations (\ref{proj3}). 

As a simple inspection of the fusion rules in Section \ref{SectionFusAlg} reveals,
the $2pp'$ ${\cal W}$-projective representations generate a closed fusion subalgebra
of (\ref{Wfus}), naturally denoted by
\be
 \big\langle\ketw{\kappa p,p'},\ketw{\R_{\kappa p,p'}^{a,0}},\ketw{\R_{p,\kappa p'}^{0,b}},
   \ketw{\R_{\kappa p,p'}^{a,b}}\big\rangle_{p,p'}
\label{projfus}
\ee
We will comment on this fusion subalgebra in Section \ref{SectionDiscussion}.

\section{Lattice realization of ${\cal WLM}(p,p')$}
\label{SectionLattice}

In \cite{PRR0803}, we used the infinite series of logarithmic minimal lattice models ${\cal LM}(1,p')$ 
to obtain ${\cal W}$-extended fusion rules applicable in the extended pictures ${\cal WLM}(1,p')$.
A crucial ingredient was the construction of a ${\cal W}$-invariant identity
representation $\ketw{1,1}$ defined as the infinite limit of a triple fusion of 
Virasoro-irreducible Kac representations in ${\cal LM}(1,p')$.
On the other hand, as indicated above and further discussed in Section \ref{SectionDiscussion},
there is no obvious natural candidate for an identity in the lattice realization of 
${\cal WLM}(p,p')$ for $p>1$. As in the case of the ${\cal W}$-extended picture of
critical percolation ${\cal WLM}(2,3)$ \cite{RP0804},
it nevertheless turns out fruitful to adopt the use of 
infinite limits of triple fusions of Virasoro-irreducible Kac representations.
This also allows us to identify the various ${\cal W}$-representations with suitable limits of
Yang-Baxter integrable boundary conditions on the lattice.
Firmly based on the lattice-realization of the fundamental fusion algebra of ${\cal LM}(p,p')$, 
our fusion prescription for ${\cal WLM}(p,p')$ yields a commutative and associative fusion algebra.
The analysis of this fusion algebra is greatly simplified by separating into
`horizontal and vertical components', much akin to the situation for the Virasoro picture
of ${\cal LM}(p,p')$ in \cite{RP0706,RP0707}.
With $p$ and $p'$ unspecified, the two components have equivalent properties
whose details only depend on the parities of $p$ and $p'$. We may therefore
focus on the horizontal component and consider it for both possible parities
knowing that the results can be translated straightforwardly to the vertical component.
Once the two components are understood, we will describe how to merge them in order
to construct the complete fusion algebra.
Since the two characterizing parameters $p$ and $p'$ are coprime, we have two
distinctively different situations, namely $p$ and $p'$ both odd or $p$ and $p'$ of different parities.

\subsection{Horizontal component}

Working in the {\em fundamental} fusion algebra of the logarithmic minimal model
${\cal LM}(p,p')$, as opposed to the lesser-understood but larger {\em full\/} fusion algebra
\cite{RP0706,RP0707}, the only horizontal Kac representations at our 
disposal\footnote{Strictly speaking, the first and finite set for $p=2$ is not generated by 
repeated fusions of the fundamental Kac representation $(2,1)$ even though it is present in 
the fundamental fusion algebra generated by repeated fusions of {\em both} of the 
fundamental representations $(2,1)$ and $(1,2)$.} 
are $\{(a,1);\ a\in\mathbb{Z}_{1,p-1}\}$ and $\{(kp,1);\ k\in\mathbb{N}\}$. 
It is noted that the second and infinite set consists of Virasoro-irreducible representations only.
There are many possible triple fusions to consider.
We find it useful to introduce $\ketw{\Ec,1}$ as the limit
\be
 \ketw{\Ec,1}:=\lim_{n\to\infty}(2np,1)^{\otimes3}
\label{E1}
\ee
and $\ketw{\Oc,1}$ as 
\be
 \ketw{\Oc,1}:=\ \frac{1}{2p}(2p,1)\otimes\ketw{\Ec,1}
\label{O1}
\ee
Here the notation $\Ec$ refers to ``even" while it refers to ``embedding" in (\ref{E}). 
This dual role should not cause confusion. 
For $p$ odd, the expression (\ref{O1}) coincides with the alternative limit
\be
 \ketw{\Oc,1}\ =\lim_{n\to\infty}((2n-1)p,1)^{\otimes3}
\label{O1podd}
\ee
whereas for $p$ even, the two limits themselves coincide.
These limits of Virasoro fusions decompose in terms of Virasoro-indecomposable representations. 
However, the technical part of the further analysis depends on the parity of $p$ so we 
initially consider the two parities separately. 
This distinction is not conceptually relevant, and as we will see, 
the main results indeed have a very simple dependence on the parity.

\subsubsection{$p$ even}
\label{sectionpeven}

In this subsection \ref{sectionpeven}, we let $p$ be even and find
\bea
 \ketw{\Ec,1}&=&\bigoplus_{\al}^{p-2}(p-\al)
   \Big\{\bigoplus_{k\in\mathbb{N}}k\R_{kp,1}^{\al,0}\Big\}\nn
  &=&\bigoplus_{\al}^{p-2}(p-\al)\Big\{
   \bigoplus_{k\in\mathbb{N}}(2k-1)\R_{(2k-1)p,1}^{\al,0}\oplus
      \bigoplus_{k\in\mathbb{N}}2k\R_{2kp,1}^{\al,0}\Big\}
\eea
where we recall the convenient notation (\ref{R00}), and
\bea
 \ketw{\Oc,1}&=&\bigoplus_{\al}^{p-1}(p-\al)
   \Big\{\bigoplus_{k\in\mathbb{N}}k\R_{kp,1}^{\al,0}\Big\}\nn
  &=&\bigoplus_{\al}^{p-1}(p-\al)\Big\{
   \bigoplus_{k\in\mathbb{N}}(2k-1)\R_{(2k-1)p,1}^{\al,0}
      \oplus\bigoplus_{k\in\mathbb{N}}2k\R_{2kp,1}^{\al,0}\Big\}
\label{E1C1peven}
\eea
Since $p$ is even and therefore greater than 1, both of these decompositions 
in terms of Virasoro-indecomposable representations are non-trivial.
Our next task is to disentangle these results and write them in terms of the 
${\cal W}$-indecomposable representations (\ref{r1}) and (\ref{r2}).
Following \cite{RP0804}, we observe that the part of $\ketw{\Ec,1}$ with $\al=0$
corresponds to a direct sum of Virasoro-irreducible representations {\em not} taking part
in any indecomposable combination. 
By selection of link states in the lattice description, we can thus 
project onto the set $\{\R_{(2k-1)p,1}^{0,0}\equiv((2k-1)p,1);\ k\in\mathbb{N}\}$ or
$\{\R_{2kp,1}^{0,0}\equiv(2kp,1);\ k\in\mathbb{N}\}$ separately, thereby allowing us to single out
the two infinite direct sums
\be
 \ketw{\kappa p,1}\ =\ \bigoplus_{k\in\mathbb{N}}(2k-2+\kappa)((2k-2+\kappa)p,1)
\label{kp1}
\ee
where $\kappa\in\mathbb{Z}_{1,2}$.
Asserting that these expressions indeed correspond to ${\cal W}$-indecomposable representations,
we have thus identified the latter with limits of Yang-Baxter integrable boundary 
conditions on the lattice accompanied by specific selections of link states.
Since the participating Virasoro representations all are of rank 1, 
the ${\cal W}$-indecomposable representation $\ketw{\kappa p,1}$ itself is of rank 1.

Having identified $\ketw{\kappa p,1}$ for each $\kappa\in\mathbb{Z}_{1,2}$, we now define the
${\cal W}$-indecomposable rank-2 representation
\be
 \ketw{\R_{\kappa p,1}^{1,0}}:=\ (2,1)\otimes\ketw{\kappa p,1}
   \ =\ \bigoplus_{k\in\mathbb{N}}(2k-2+\kappa)\R_{(2k-2+\kappa)p,1}^{1,0}
\label{R2121}
\ee 
This would complete the disentanglement of (\ref{E1}) for $p=2$. For $p>2$, we continue by
decomposing the more general fusion
\be
 (r,1)\otimes\ketw{\kappa p,1}\ =\ 
   \bigoplus_{\al}^{r-1}
      \Big\{\bigoplus_{k\in\mathbb{N}}(2k-2+\kappa)\R_{(2k-2+\kappa)p,1}^{\al,0}\Big\}
\ee
where we recall the direct-sum convention (\ref{sum2}).
For $r=3$, the sum over $\al$ involves two terms of which the one for
$\al=0$ is recognized as $\ketw{\kappa p,1}$. The other term is subsequently
identified with
\be
 \ketw{\R_{\kappa p,1}^{2,0}}
   \ =\ \bigoplus_{k\in\mathbb{N}}(2k-2+\kappa)\R_{(2k-2+\kappa)p,1}^{2,0}
\ee
For $r=4$, the sum over $\al$ also involves two terms of which the one for
$\al=1$ is recognized as $\ketw{\R_{\kappa p,1}^{1,0}}$. The other term is subsequently
identified with
\be
 \ketw{\R_{\kappa p,1}^{3,0}}
   \ =\ \bigoplus_{k\in\mathbb{N}}(2k-2+\kappa)\R_{(2k-2+\kappa)p,1}^{3,0}
\ee
For $r=5$, the sum over $\al$ involves three terms of which the ones for
$\al=0$ and $\al=2$ are recognized as $\ketw{\kappa p,1}$ and
$\ketw{\R_{\kappa p,1}^{2,0}}$, respectively. We can subsequently identify the remaining term
with $\ketw{\R_{\kappa p,1}^{4,0}}$. It is now clear how a bootstrapping
procedure, as $r$ increases to its greatest possible value $p$, allows
us to identify 
\be
 \ketw{\R_{\kappa p,1}^{a,0}}
   \ =\ \bigoplus_{k\in\mathbb{N}}(2k-2+\kappa)\R_{(2k-2+\kappa)p,1}^{a,0}
\label{Rkp1}
\ee
for all $a\in\mathbb{Z}_{1,p-1}$.
We assert that these representations are ${\cal W}$-indecomposable
and note that they all are of rank 2.
In conclusion, we have found that the representations $\ketw{\Ec,1}$ and $\ketw{\Oc,1}$ are
${\cal W}$-decomposable as they can be written
as the following direct sums of ${\cal W}$-indecomposable representations
\bea
 \ketw{\Ec,1}
  &=&\bigoplus_{\al}^{p-2}(p-\al)\Big\{
   \ketw{\R_{p,1}^{\al,0}}\oplus\ketw{\R_{2p,1}^{\al,0}}\Big\}\nn
 \ketw{\Oc,1}
  &=&\bigoplus_{\al}^{p-1}(p-\al)\Big\{
    \ketw{\R_{p,1}^{\al,0}}\oplus\ketw{\R_{2p,1}^{\al,0}}\Big\}
\label{E1even}
\eea

Of interest in their own right, but also of importance for the evaluation of fusion products below, 
we find that the ${\cal W}$-indecomposable representations (\ref{kp1}) and (\ref{Rkp1}) 
have the `stability properties'
\bea
 \!((2n-2+\kappa)p,1)\otimes\ketw{\kappa' p,1}&=&(2n-2+\kappa)\Big\{\bigoplus_{\al}^{p-1}
      \ketw{\R_{(\kappa\cdot\kappa')p,1}^{\al,0}}\Big\} \nn
 \!\!\!((2n-2+\kappa)p,1)\otimes\ketw{\R_{\kappa' p,1}^{a,0}}
   &=&2(2n-2+\kappa)\Big\{\bigoplus_{\al}^{a-1}
   \ketw{\R_{(2\cdot\kappa\cdot\kappa')p,1}^{\al,0}}
     \oplus\bigoplus_{\al}^{p-a-1}
   \!\ketw{\R_{(\kappa\cdot\kappa')p,1}^{\al,0}}\Big\}
\label{horstab1}
\eea
and
\bea
 (r,1)\otimes\ketw{\kappa p,1}&=&\bigoplus_{\al}^{r-1}
      \ketw{\R_{\kappa p,1}^{\al,0}}  \nn
 (r,1)\otimes\ketw{\R_{\kappa p,1}^{a,0}}
   &=&\!\!\!\bigoplus_{\al=|a-r|+1,\ \!{\rm by}\ \!2}^{p-|p-r-a|-1}
      \!\!\!\ketw{\R_{\kappa p,1}^{\al,0}}\oplus\!\bigoplus_{\al}^{r-a-1}\!2\ketw{\R_{\kappa p,1}^{\al,0}}
     \oplus\!\!\bigoplus_{\al}^{a+r-p-1}
        \!\!\!2\ketw{\R_{(2\cdot\kappa)p,1}^{\al,0}}
\label{horstab2}
\eea
We note that the two expressions for $(p,1)\otimes\ketw{\kappa p,1}$ (and likewise for
$(p,1)\otimes\ketw{\R_{\kappa p,1}^{a,0}}$) appearing in (\ref{horstab1}) and (\ref{horstab2}), 
respectively, agree. It also follows that the ${\cal W}$-representations $\ketw{\Ec,1}$ and 
$\ketw{\Oc,1}$ are related by the remarkably simple stability properties
\be
 (2np,1)\otimes\ketw{\Ec,1}\ =\ 2np\ketw{\Oc,1},\qquad\quad 
 (2np,1)\otimes\ketw{\Oc,1}\ =\ 2np\ketw{\Ec,1}
\label{2np}
\ee
and
\be
 (r,1)\otimes\ketw{\Ec,1}
   \ =\ \begin{cases}r\ketw{\Oc,1},\quad &r\ \mathrm{even} \\ 
     r\ketw{\Ec,1},\quad &r\ \mathrm{odd} \end{cases}
   \qquad\quad
 (r,1)\otimes\ketw{\Oc,1}
   \ =\ \begin{cases}r\ketw{\Ec,1},\quad &r\ \mathrm{even} \\ 
     r\ketw{\Oc,1},\quad &r\ \mathrm{odd} \end{cases}
\label{r1EO}
\ee
As we will see in the following, there are many more such properties, but this list suffices for now.

{}From the lattice, we define the ${\cal W}$-extended fusion product $\hat\otimes$ by
\be
 \ketw{\Ec,1}\fus\ketw{A}:=\ 
  \lim_{n\to\infty}\Big(\frac{1}{2n}\Big)^3(2np,1)^{\otimes3}\otimes\ketw{A}
\label{E1A}
\ee
First, we consider the two cases $\ketw{A}=\ketw{\kappa p,1}$ 
where $\kappa\in\mathbb{Z}_{1,2}$ and find
\bea
  \ketw{\Ec,1}\otimes\ketw{\kappa p,1}&=&\bigoplus_{\al}^{p-2}(p-\al)\Big\{
   \ketw{\R_{p,1}^{\al,0}}\oplus\ketw{\R_{2p,1}^{\al,0}}\Big\}\fus\ketw{\kappa p,1}\nn
    &=&\lim_{n\to\infty}\Big(\frac{1}{2n}\Big)^3(2np,1)^{\otimes3}\otimes\ketw{\kappa p,1}\nn
  &=&\lim_{n\to\infty}\Big(\frac{1}{2n}\Big)^2(2np,1)^{\otimes2}\otimes 
    \Big(\bigoplus_{\al}^{p-1}\ketw{\R_{(2\cdot\kappa)p,1}^{\al,0}}\Big)\nn
  &=&\lim_{n\to\infty}\Big(\frac{1}{2n}\Big)(2np,1)\otimes
    \Big(\bigoplus_{\al}^{p-2}(p-\al)\big\{\ketw{\R_{p,1}^{\al,0}}\oplus
      \ketw{\R_{2p,1}^{\al,0}}\big\}\Big)\nn
   &=&\bigoplus_{\al}^{p-1}p(p-\al)\Big\{
    \ketw{\R_{p,1}^{\al,0}}\oplus\ketw{\R_{2p,1}^{\al,0}}\Big\}\nn
  &=&p\ketw{\Oc,1}
\label{214121}
\eea
which is seen to be {\em independent} of $\kappa$.
We are still faced with the task of disentangling these results since the
identification of the individual fusions such as $\ketw{p,1}\fus\ketw{p,1}$ is ambiguous at this point. 
To this end, we use (\ref{kpkpRkpkp}) to deduce that the decomposition of the fusion 
$\ketw{\kappa p,1}\fus\ketw{\kappa' p,1}$ only
involves representations of the form $\ketw{\R_{(\kappa\cdot\kappa')p,1}^{\al,0}}$ with $\al$ odd
and that these ${\cal W}$-indecomposable rank-2 representations 
only appear there in multiples of the combination
$\bigoplus_{\al}^{p-1}\ketw{\R_{(\kappa\cdot\kappa')p,1}^{\al,0}}$.
It also follows from (\ref{kpkpRkpkp}) that, in the fusion $\ketw{\Ec,1}\fus\ketw{p,1}$,
the ${\cal W}$-indecomposable rank-2 representation $\ketw{\R_{p,1}^{p-1,0}}$ is only
produced by $\ketw{p,1}\fus\ketw{p,1}$. Since $\ketw{p,1}$ appears with multiplicity $p$ in the 
decomposition of $\ketw{\Ec,1}$, and $\ketw{\R_{p,1}^{p-1,0}}$ appears with multiplicity 
$p$ in the fusion $\ketw{\Ec,1}\fus\ketw{p,1}$, it follows that $\ketw{\R_{p,1}^{p-1,0}}$ is 
produced with multiplicity $1$ in the fusion $\ketw{p,1}\fus\ketw{p,1}$. 
We thus conclude that
\be
 \ketw{p,1}\fus\ketw{p,1}
  \ =\ \bigoplus_{\al}^{p-1}\ketw{\R_{p,1}^{\al,0}}
\ee
We likewise find that
\be
 \ketw{p,1}\fus\ketw{2p,1}
  \ =\ \bigoplus_{\al}^{p-1}\ketw{\R_{2p,1}^{\al,0}},\qquad\quad
 \ketw{2p,1}\fus\ketw{2p,1}\ =\  \bigoplus_{\al}^{p-1}\ketw{\R_{p,1}^{\al,0}}
\ee

In order to evaluate fusions involving the ${\cal W}$-indecomposable rank-2 representations 
$\ketw{\R_{\kappa p,1}^{a,0}}$, we note that (\ref{horstab2}) implies
\be
 (a+1,1)\otimes\ketw{\kappa p,1}\ =\ \ketw{\R_{\kappa p,1}^{a,0}}
   \oplus\Big((a-1,1)\otimes\ketw{\kappa p,1}\Big)
\ee
It follows that
\be
 \ketw{\R_{\kappa p,1}^{a,0}}\fus\ketw{A}\ =\ (a+1,1)\otimes\Big(\ketw{\kappa p,1}\fus\ketw{A}\Big)
  \ominus(a-1,1)\otimes\Big(\ketw{\kappa p,1}\fus\ketw{A}\Big)
\ee
(where $\ominus$ denotes direct subtraction) which for $\ketw{A}=\ketw{\kappa' p,1}$ yields
\be
  \ketw{\R_{\kappa p,1}^{a,0}}\fus\ketw{\kappa' p,1}
  \ =\ \bigoplus_{\al}^{p-a-1}
     \!2\ketw{\R_{(\kappa\cdot\kappa')p,1}^{\al,0}}
  \oplus\bigoplus_{\al}^{a-1}
     2\ketw{\R_{(2\cdot\kappa\cdot\kappa')p,1}^{\al,0}} 
\ee
By re-cycling these results with $\ketw{A}=\ketw{\R_{\kappa' p,1}^{a',0}}$, we finally obtain
\bea
 \ketw{\R_{\kappa p,1}^{a,0}}\fus\ketw{\R_{\kappa' p,1}^{a',0}}
   &=&\bigoplus_{\al}^{p-|a-a'|-1}
     \!\!2\ketw{\R_{(\kappa\cdot\kappa')p,1}^{\al,0}}
     \oplus\bigoplus_{\al}^{|p-a-a'|-1}
     \!\!2\ketw{\R_{(\kappa\cdot\kappa')p,1}^{\al,0}}\nn
   &\oplus&\bigoplus_{\al}^{p-|p-a-a'|-1}
     \!\!\!2\ketw{\R_{(2\cdot\kappa\cdot\kappa')p,1}^{\al,0}}
     \oplus\bigoplus_{\al}^{|a-a'|-1}
     \!2\ketw{\R_{(2\cdot\kappa\cdot\kappa')p,1}^{\al,0}}
\eea

\subsubsection{$p$ odd}
\label{sectionpodd}

In this subsection \ref{sectionpodd}, we let $p$ be odd and find
\be
 \ketw{\Ec,1}\ =\ \bigoplus_{\al}^{p-1}(p-\al)
   \Big\{\bigoplus_{k\in\mathbb{N}}2k\R_{2kp,1}^{\al,0}\Big\}
  \oplus\bigoplus_{\al}^{p-2}(p-\al)
    \Big\{\bigoplus_{k\in\mathbb{N}}(2k-1)\R_{(2k-1)p,1}^{\al,0}\Big\}
\label{Eodd}
\ee
and
\be
 \ketw{\Oc,1}\ =\ \bigoplus_{\al}^{p-1}(p-\al)
    \Big\{\bigoplus_{k\in\mathbb{N}}(2k-1)\R_{(2k-1)p,1}^{\al,0}\Big\}
  \oplus\bigoplus_{\al}^{p-2}(p-\al)
    \Big\{\bigoplus_{k\in\mathbb{N}}2k\R_{2kp,1}^{\al,0}\Big\}
\label{Oodd}
\ee
Decompositions of these results in terms of ${\cal W}$-indecomposable representations are
obtained by mimicking the disentangling procedure employed above for $p$ even. 
That is, by appropriately selecting the link states in the lattice description, we first isolate
the terms corresponding to $\al=0$ in (\ref{Eodd}) and (\ref{Oodd})
\be
 \ketw{2p,1}\ =\ \bigoplus_{k\in\mathbb{N}}2k(2kp,1),\qquad\quad
 \ketw{p,1}\ =\ \bigoplus_{k\in\mathbb{N}}(2k-1)((2k-1)p,1)
\label{2p1p1}
\ee
Having identified these ${\cal W}$-indecomposable rank-1 representations,
we apply the bootstrapping procedure where the ${\cal W}$-indecomposable 
rank-2 representations
\be
 \ketw{\R_{\kappa p,1}^{a,0}}\ =\ \bigoplus_{k\in\mathbb{N}}(2k-2+\kappa)\R_{(2k-2+\kappa)p,1}^{a,0}
\label{Ra0odd}
\ee
are identified one by one as $a$ increases from 1 to $p-1$ in
\be
 (a+1,1)\otimes\ketw{\kappa p,1}\ =\ 
   \bigoplus_{\al}^{a}
      \Big\{\bigoplus_{k\in\mathbb{N}}(2k-2+\kappa)\R_{(2k-2+\kappa)p,1}^{\al,0}\Big\}
\ee
Asserting that the representations (\ref{2p1p1}) and (\ref{Ra0odd}) indeed are 
${\cal W}$-indecomposable, we see that the decompositions of $\ketw{\Ec,1}$ and $\ketw{\Oc,1}$
in terms of ${\cal W}$-indecomposable representations read
\bea
 \ketw{\Ec,1}&=&\bigoplus_{\al}^{p-2}(p-\al)\ketw{\R_{p,1}^{\al,0}}
   \oplus\bigoplus_{\al}^{p-1}(p-\al)
    \ketw{\R_{2p,1}^{\al,0}}  \nn
 \ketw{\Oc,1}&=&\bigoplus_{\al}^{p-1}(p-\al)
    \ketw{\R_{p,1}^{\al,0}}
  \oplus\bigoplus_{\al}^{p-2}(p-\al)
    \ketw{\R_{2p,1}^{\al,0}}
\label{E1odd}
\eea
We note that the horizontal ${\cal W}$-indecomposable representations have stability properties
which our notation (\ref{sum2}) allows us to write in the exact same way as for $p$ even,
namely (\ref{horstab1}) and (\ref{horstab2}).
Likewise, the representations $\ketw{\Ec,1}$ and $\ketw{\Oc,1}$ have the same simple stability
properties (\ref{2np}) and (\ref{r1EO}) as for $p$ even.
Fusion is naturally defined as for $p$ even (\ref{E1A}). Thus following the derivation of the fusion
rules for $p$ even, but now based on the stability properties just listed, we obtain the
fusion rules for $p$ odd. They are given in the summary below.

\subsubsection{General $p$}
\label{SectionGeneralp}

In summary, valid for both parities of $p$, we have determined the horizontal fusion rules  
\bea
 \ketw{\kappa p,1}\fus\ketw{\kappa' p,1}
  &=&\bigoplus_{\al}^{p-1}\ketw{\R_{(\kappa\cdot\kappa')p,1}^{\al,0}} \nn
 \ketw{\kappa p,1}\fus\ketw{\R_{\kappa' p,1}^{a,0}}
  &=&\bigoplus_{\al}^{p-a-1}
     2\ketw{\R_{(\kappa\cdot\kappa')p,1}^{\al,0}}
  \oplus\bigoplus_{\al}^{a-1}
     2\ketw{\R_{(2\cdot\kappa\cdot\kappa')p,1}^{\al,0}} \nn
 \ketw{\R_{\kappa p,1}^{a,0}}\fus\ketw{\R_{\kappa' p,1}^{a',0}}
  &=&\bigoplus_{\al}^{p-|a-a'|-1}
     \!\!2\ketw{\R_{(\kappa\cdot\kappa')p,1}^{\al,0}}
     \oplus\bigoplus_{\al}^{|p-a-a'|-1}
     \!\!2\ketw{\R_{(\kappa\cdot\kappa')p,1}^{\al,0}}\nn
  &\oplus&\bigoplus_{\al}^{p-|p-a-a'|-1}
     \!\!\!2\ketw{\R_{(2\cdot\kappa\cdot\kappa')p,1}^{\al,0}}
     \oplus\bigoplus_{\al}^{|a-a'|-1}
     \!2\ketw{\R_{(2\cdot\kappa\cdot\kappa')p,1}^{\al,0}}
\eea
governing the $2p$-dimensional closed fusion (sub)algebra
\be
 \big\langle\ketw{\kappa p,1},\ketw{\R_{\kappa p,1}^{a,0}}\big\rangle_{p,p'}
\label{horpp}
\ee
We note that the two representations $\ketw{\Ec,1}$ and $\ketw{\Oc,1}$ 
form a two-dimensional subalgebra of this fusion algebra
\be
 \ketw{\Ec,1}\fus\ketw{\Ec,1}\ =\ \ketw{\Oc,1}\fus\ketw{\Oc,1}\ =\ p^3\ketw{\Oc,1},\qquad\quad 
 \ketw{\Ec,1}\fus\ketw{\Oc,1}\ =\ p^3\ketw{\Ec,1}
\label{EEeven}
\ee

\subsection{Vertical component}
\label{SectionVertical}

The vertical component is obtained from the horizontal component simply by interchanging
the first and second indices and replacing $p$ by $p'$ as in
\be
 \{\ketw{\R_{\kappa p,1}^{\al,0}};\ \al\in\mathbb{Z}_{0,p-1}\}\qquad \longrightarrow\qquad
 \{\ketw{\R_{1,\kappa p'}^{0,\beta}};\ \beta\in\mathbb{Z}_{0,p'-1}\} 
\ee
where the vertical ${\cal W}$-representations decompose in terms of Virasoro-indecomposable
representations as
\bea
 \ketw{1,\kappa p'}&=&\bigoplus_{k\in\mathbb{N}}(2k-2+\kappa)(1,(2k-2+\kappa)p')\nn
 \ketw{\R_{1,\kappa p'}^{0,b}}
   &=&\bigoplus_{k\in\mathbb{N}}(2k-2+\kappa)\R_{1,(2k-2+\kappa)p'}^{0,b}
\label{1kp}
\eea
The $2p'$-dimensional vertical fusion (sub)algebra
\be
 \big\langle\ketw{1,\kappa p'},\ketw{\R_{1,\kappa p'}^{0,b}}\big\rangle_{p,p'}
\label{verpp}
\ee
is thus governed by the fusion rules
\bea
 \ketw{1,\kappa p'}\fus\ketw{1,\kappa' p'}
  &=&\bigoplus_{\beta}^{p'-1}\ketw{\R_{1,(\kappa\cdot\kappa')p'}^{0,\beta}} \nn
 \ketw{1,\kappa p'}\fus\ketw{\R_{1,\kappa' p'}^{0,b}}
  &=&\bigoplus_{\beta}^{p'-b-1}
     2\ketw{\R_{1,(\kappa\cdot\kappa')p'}^{0,\beta}}
  \oplus\bigoplus_{\beta}^{b-1}
     2\ketw{\R_{1,(2\cdot\kappa\cdot\kappa')p'}^{0,\beta}}  \nn
 \ketw{\R_{1,\kappa p'}^{0,b}}\fus\ketw{\R_{1,\kappa' p'}^{0,b'}}
  &=&\bigoplus_{\beta}^{p'-|b-b'|-1}
     \!\!2\ketw{\R_{1,(\kappa\cdot\kappa')p'}^{0,\beta}}
     \oplus\bigoplus_{\beta}^{|p'-b-b'|-1}
     \!\!2\ketw{\R_{1,(\kappa\cdot\kappa')p'}^{0,\beta}}\nn
   &\oplus&\bigoplus_{\beta}^{p'-|p'-b-b'|-1}
     \!\!\!2\ketw{\R_{1,(2\cdot\kappa\cdot\kappa')p'}^{0,\beta}}
     \oplus\bigoplus_{\beta}^{|b-b'|-1}
     \!2\ketw{\R_{1,(2\cdot\kappa\cdot\kappa')p'}^{0,\beta}}
\eea

Of course, these results can be obtained `directly' from the lattice
by introducing the vertical representation $\ketw{1,\Ec}$ as the limit
\be
 \ketw{1,\Ec}:=\lim_{n\to\infty}(1,2np')^{\otimes3}
\label{1E}
\ee
and its companion $\ketw{1,\Oc}$ by
\be
 \ketw{1,\Oc}:=\ \frac{1}{2p'}(1,2p')\otimes\ketw{1,\Ec}
\label{1O}
\ee
They have the stability properties
\be
 (1,2np')\otimes\ketw{1,\Ec}\ =\ 2np'\ketw{1,\Oc},\qquad\quad 
 (1,2np')\otimes\ketw{1,\Oc}\ =\ 2np'\ketw{1,\Ec}
\label{2npv}
\ee
and
\be
 \ketw{1,\Ec}\otimes(1,s)
   \ =\ \begin{cases}s\ketw{1,\Oc},\quad &s\ \mathrm{even} \\ 
     s\ketw{1,\Ec},\quad &s\ \mathrm{odd} \end{cases}
   \qquad\quad
 \ketw{1,\Oc}\otimes(1,s)
   \ =\ \begin{cases}s\ketw{1,\Ec},\quad &s\ \mathrm{even} \\ 
     s\ketw{1,\Oc},\quad &s\ \mathrm{odd} \end{cases}
\label{EO1s}
\ee
while the ${\cal W}$-indecomposable representations have the stability properties
\bea
 \!\!\!\!\!\!\!(1,(2n-2+\kappa)p')\otimes\ketw{1,\kappa' p'}&=&(2n-2+\kappa)\Big\{\bigoplus_{\beta}^{p'-1}
      \ketw{\R_{1,(\kappa\cdot\kappa')p'}^{0,\beta}}\Big\} \nn
  \!\!\!\!\!\!\!\!\!(1,(2n-2+\kappa)p')\otimes\ketw{\R_{1,\kappa' p'}^{0,b}}
   &=&2(2n-2+\kappa)\Big\{\bigoplus_{\beta}^{b-1}
   \ketw{\R_{1,(2\cdot\kappa\cdot\kappa')p'}^{0,\beta}}
     \oplus\!\bigoplus_{\beta}^{p'-b-1}
   \!\ketw{\R_{1,(\kappa\cdot\kappa')p'}^{0,\beta}}\Big\}
\label{verstab1}
\eea
and
\bea
 (1,s)\otimes\ketw{1,\kappa p'}&=&\bigoplus_{\beta}^{s-1}
      \ketw{\R_{1,\kappa p'}^{0,\beta}}  \nn
 (1,s)\otimes\ketw{\R_{1,\kappa p'}^{0,b}}
   &=&\!\!\!\bigoplus_{\beta=|b-s|+1,\ \!{\rm by}\ \!2}^{p'-|p'-s-b|-1}
      \!\!\!\ketw{\R_{1,\kappa p'}^{0,\beta}}
   \oplus\!\bigoplus_{\beta}^{s-b-1}\!2\ketw{\R_{1,\kappa p'}^{0,\beta}}
     \oplus\!\!\!\bigoplus_{\beta}^{b+s-p'-1}
        \!\!\!\!2\ketw{\R_{1,(2\cdot\kappa)p'}^{0,\beta}}
\label{verstab2}
\eea
In accordance with the definition (\ref{E1A}), the individual vertical fusions then follow from 
appropriately disentangling the result of evaluating
\be
 \ketw{A}\fus\ketw{1,\Ec}\ =\ 
  \lim_{n\to\infty}\Big(\frac{1}{2n}\Big)^3\ketw{A}\otimes(1,2np')^{\otimes3}
\label{A1E}
\ee
For $p'$ even, the decompositions of $\ketw{1,\Ec}$ and $\ketw{1,\Oc}$ in terms of 
${\cal W}$-indecomposable representations read
\bea
 \ketw{1,\Ec}
  &=&\bigoplus_{\beta}^{p'-2}(p'-\beta)\Big\{
   \ketw{\R_{1,p'}^{0,\beta}}\oplus\ketw{\R_{1,2p'}^{0,\beta}}\Big\}\nn
 \ketw{1,\Oc}
  &=&\bigoplus_{\beta}^{p'-1}(p'-\beta)\Big\{
    \ketw{\R_{1,p'}^{0,\beta}}\oplus\ketw{\R_{1,2p'}^{0,\beta}}\Big\}
\label{1Eeven}
\eea
while for $p'$ odd they read
\bea
 \ketw{1,\Ec}&=&\bigoplus_{\beta}^{p'-2}(p'-\beta)\ketw{\R_{1,p'}^{0,\beta}}
   \oplus\bigoplus_{\beta}^{p'-1}(p'-\beta)\ketw{\R_{1,2p'}^{0,\beta}}   \nn
 \ketw{1,\Oc}&=&\bigoplus_{\beta}^{p'-1}(p'-\beta)\ketw{\R_{1,p'}^{0,\beta}}
  \oplus\bigoplus_{\beta}^{p'-2}(p'-\beta)  \ketw{\R_{1,2p'}^{0,\beta}}
\label{1Eodd}
\eea
The two-dimensional fusion subalgebra generated by $\ketw{1,\Ec}$ and $\ketw{1,\Oc}$ 
is governed by the fusion rules
\be
 \ketw{1,\Ec}\fus\ketw{1,\Ec}\ =\ \ketw{1,\Oc}\fus\ketw{1,\Oc}\ =\ (p')^3\ketw{1,\Oc},\qquad\quad 
 \ketw{1,\Ec}\fus\ketw{1,\Oc}\ =\ (p')^3\ketw{1,\Ec}
\label{1E1E}
\ee

\subsection{Horizontal and vertical components combined}

Here we describe the merge of the horizontal and vertical components by completing the
set of $N_{\mathrm{ind}}(p,p')$ ${\cal W}$-representations announced in (\ref{Ntotal}). 
The derivation of the ensuing fusion algebra (\ref{Wfus}) is discussed in Section \ref{SectionFusion}.

\subsubsection{Representation content}
\label{SectionRepContent}

New representations are constructed by fusing the horizontal representations
above by the simple vertical (Virasoro-indecomposable) Kac representations $(1,s)$.  
For $s\in\mathbb{Z}_{2,p'}$, we thus define the ${\cal W}$-indecomposable rank-1 representations
\be
 \ketw{\kappa p,s}:=\ \ketw{\kappa p,1}\otimes(1,s)
   \ =\ \bigoplus_{k\in\mathbb{N}}(2k-2+\kappa)((2k-2+\kappa)p,s)
\label{2s}
\ee
and the ${\cal W}$-indecomposable rank-2 representations
\be
 \ketw{\R_{\kappa p,s}^{a,0}}:=\ \ketw{\R_{\kappa p,1}^{a,0}}\otimes(1,s)
   \ =\ \bigoplus_{k\in\mathbb{N}}(2k-2+\kappa)\R_{(2k-2+\kappa)p,s}^{a,0}
\label{R2s}
\ee
For $s=1$, these identities are valid but do not constitute definitions.
We also define ${\cal W}$-indecomposable rank-3 representations by fusing 
the ${\cal W}$-indecomposable rank-2 representations with
vertical Virasoro-indecomposable representations of rank 2
\be
 \ketw{\R_{\kappa p,p'}^{a,b}}:=\ \ketw{\R_{\kappa p,1}^{a,0}}\otimes\R_{1,p'}^{0,b}
   \ =\ \bigoplus_{k\in\mathbb{N}}(2k-2+\kappa)\R_{(2k-2+\kappa)p,p'}^{a,b}
\ee
We could just as well have fused the {\em vertical} ${\cal W}$-representations
by {\em horizontal} Virasoro-indecomposable representations.
For $r\in\mathbb{Z}_{2,p}$, this yields the ${\cal W}$-indecomposable rank-1 representations
\be
 \ketw{r,\kappa p'}:=\ (r,1)\otimes\ketw{1,\kappa p'}
   \ =\ \bigoplus_{k\in\mathbb{N}}(2k-2+\kappa)(r,(2k-2+\kappa)p')
\ee
the ${\cal W}$-indecomposable rank-2 representations
\be
 \ketw{\R_{r,\kappa p'}^{0,b}}:=\ (r,1)\otimes\ketw{\R_{1,\kappa p'}^{0,b}}
   \ =\ \bigoplus_{k\in\mathbb{N}}(2k-2+\kappa)\R_{r,(2k-2+\kappa)p'}^{0,b}
\label{Rr2}
\ee
and the ${\cal W}$-indecomposable rank-3 representations
\be
 \ketw{\R_{p,\kappa p'}^{a,b}}:=\ \R_{p,1}^{a,0}\otimes\ketw{\R_{\kappa p'}^{0,b}}
   \ =\ \bigoplus_{k\in\mathbb{N}}(2k-2+\kappa)\R_{p,(2k-2+\kappa)p'}^{a,b}
\ee
The identities $\R_{kp,k'p'}^{\al,\beta}\equiv\R_{k'p,kp'}^{\al,\beta}$ between Virasoro representations
imply the ${\cal W}$-representation identities
\be
 \ketw{2p,p'}\ \equiv\ \ketw{p,2p'},\qquad\quad\ketw{\R_{2p,p'}^{a,b}}\ \equiv\ \ketw{\R_{p,2p'}^{a,b}}
\ee
For convenience of notation, we also introduce
\be
  \ketw{\R_{2p,2p'}^{\al,\beta}}:=\ \frac{1}{2}\R_{2p,1}^{\al,0}\otimes\ketw{\R_{1,2p'}^{0,\beta}}
    \ =\ \bigoplus_{k\in\mathbb{N}}k\R_{2p,2kp'}^{\al,\beta}
    \ =\ \bigoplus_{k\in\mathbb{N}}(2k-1)\ \R_{p,(2k-1)p'}^{\al,\beta}
    \ =\ \ketw{\R_{p,p'}^{\al,\beta}}
\ee
Compactly, our notation allows us to write 
\be
 \ketw{\R_{(\kappa\cdot\kappa')p,p'}^{\al,\beta}}
  \ \equiv\ \ketw{\R_{\kappa p,\kappa' p'}^{\al,\beta}}
  \ \equiv\ \ketw{\R_{p,(\kappa\cdot\kappa')p'}^{\al,\beta}} 
\ee
and
\be
 \frac{1}{\kappa'}\ketw{\R_{\kappa p,1}^{\al,0}}\otimes\R_{1,\kappa' p'}^{0,\beta}
  \ =\ \ketw{\R_{\kappa p,\kappa' p'}^{\al,\beta}}
  \ =\ \frac{1}{\kappa}\R_{\kappa p,1}^{\al,0}\otimes\ketw{\R_{1,\kappa' p'}^{0,\beta}}
\label{frackappa}
\ee
Having ventured into the bulk part of the Kac table, we note the stability properties
\bea
 \ketw{\R_{\kappa p,1}^{\al,0}}\otimes(1,(2n-2+\kappa')p')
  &=&(2n-2+\kappa')\ketw{\R_{(\kappa\cdot\kappa')p,p'}^{\al,0}}\nn
 ((2n-2+\kappa)p,1)\otimes\ketw{\R_{1,\kappa' p'}^{0,\beta}}
  &=&(2n-2+\kappa)\ketw{\R_{p,(\kappa\cdot\kappa')p'}^{0,\beta}}
\eea
Combining all the ${\cal W}$-indecomposable representations discussed so far in this Section
\ref{SectionLattice}, we arrive at the classification (\ref{r1}), (\ref{r2}) and (\ref{r3}).

\subsubsection{Some linear relations}
\label{SectionLinRel}

Here we list some intriguing linear relations involving
the horizontal representations $\ketw{\Ec,1}$ and $\ketw{\Oc,1}$ and the vertical
representations $\ketw{1,\Ec}$ and $\ketw{1,\Oc}$.
These linear relations will resurface when discussing fusion subalgebras in Section \ref{SectionSub}.
For $p$ even and $p'$ odd, we find that
\bea
 \bigoplus_{\beta=0}^{p'-1}(p'-\beta)\ketw{\Ec,1}\otimes\R_{1,p'}^{0,\beta}
  &=&\bigoplus_{\al}^{p-2}(p-\al)\R_{p,1}^{\al,0}\otimes
    \Big\{\ketw{1,\Ec}\oplus\ketw{1,\Oc}\Big\}\nn
 \bigoplus_{\beta=0}^{p'-1}(p'-\beta)\ketw{\Oc,1}\otimes\R_{1,p'}^{0,\beta}
  &=&\bigoplus_{\al}^{p-1}(p-\al)\R_{p,1}^{\al,0}\otimes
    \Big\{\ketw{1,\Ec}\oplus\ketw{1,\Oc}\Big\}
\label{lin1}
\eea
where it is noted that the summations over $\beta$ are in steps of 1 while the summations over $\al$ 
are in steps of 2, cf. our convention (\ref{sum2}).
For $p$ odd and $p'$ even, we similarly have
\bea
 \bigoplus_{\beta}^{p'-2}(p'-\beta)
    \Big\{\ketw{\Ec,1}\oplus\ketw{\Oc,1}\Big\}\otimes\R_{1,p'}^{0,\beta}
  &=&\bigoplus_{\al=0}^{p-1}(p-\al)\R_{p,1}^{\al,0}\otimes\ketw{1,\Ec} \nn
 \bigoplus_{\beta}^{p'-1}(p'-\beta)
    \Big\{\ketw{\Ec,1}\oplus\ketw{\Oc,1}\Big\}\otimes\R_{1,p'}^{0,\beta}
  &=&\bigoplus_{\al=0}^{p-1}(p-\al)\R_{p,1}^{\al,0}\otimes\ketw{1,\Oc} 
\label{lin2}
\eea
where it is noted that the summations over $\al$ are in steps of 1 while the summations over 
$\beta$ are in steps of 2.
Finally, for $p$ and $p'$ both odd, we have
\bea
 &&\Big\{\bigoplus_{\beta}^{p'-2}(p'-\beta)
     \ketw{\Ec,1}\otimes\R_{1,p'}^{0,\beta}\Big\}\oplus
  \Big\{\bigoplus_{\beta}^{p'-1}(p'-\beta)\ketw{\Oc,1}\otimes\R_{1,p'}^{0,\beta}\Big\}\nn
 &&\hspace{2.1cm}\ =\ \Big\{\bigoplus_{\al}^{p-2}
    (p-\al)\R_{p,1}^{\al,0}\otimes\ketw{1,\Ec}\Big\}\oplus
  \Big\{\bigoplus_{\al}^{p-1}(p-\al)\R_{p,1}^{\al,0}\otimes\ketw{1,\Oc}\Big\}\nn
  &&\Big\{\bigoplus_{\beta}^{p'-1}(p'-\beta)
    \ketw{\Ec,1}\otimes\R_{1,p'}^{0,\beta}\Big\}\oplus
  \Big\{\bigoplus_{\beta}^{p'-2}(p'-\beta)\ketw{\Oc,1}\otimes\R_{1,p'}^{0,\beta}\Big\} \nn
 &&\hspace{2.1cm}\ =\ \Big\{\bigoplus_{\al}^{p-1}
     (p-\al)\R_{p,1}^{\al,0}\otimes\ketw{1,\Ec}\Big\}\oplus
  \Big\{\bigoplus_{\al}^{p-2}(p-\al)\R_{p,1}^{\al,0}\otimes\ketw{1,\Oc}\Big\}
\label{lin3}
\eea

\subsection{${\cal W}$-extended fusion}
\label{SectionFusion}

There are two obvious approaches to the examination of fusions between horizontal and vertical
representations, namely (\ref{E1A}) and (\ref{A1E}). Self-consistency of our fusion prescription
requires that the evaluation of a given fusion product based on (\ref{E1A}) must yield the same result
as the evaluation of the same fusion product based on (\ref{A1E}), when both methods are
applicable. Since the parameters 
$p$ and $p'$ are coprime, at least one of them must be odd. Without loss of generality,
we assume that $p$ is odd and initially use (\ref{E1A}). Had we instead
assumed that $p'$ is odd, we would initially use (\ref{A1E}). 
We will subsequently address the question of self-consistency. We thus consider
\bea
 \ketw{\Ec,1}\fus\ketw{\R_{1,\kappa' p'}^{0,\beta}}
  &=&\lim_{n\to\infty}\Big(\frac{1}{2n}\Big)^3(2np,1)^{\otimes 3}\otimes\ketw{\R_{1,\kappa' p'}^{0,\beta}}\nn
  &=&\lim_{n\to\infty}\Big(\frac{1}{2n}\Big)^2(2np,1)^{\otimes 2}
    \otimes\ketw{\R_{p,(2\cdot\kappa')p'}^{0,\beta}}\nn
  &=&\lim_{n\to\infty}\Big(\frac{1}{2n}\Big)^2(p,1)\otimes(2np,1)^{\otimes 2}
    \otimes\ketw{\R_{1,(2\cdot\kappa')p'}^{0,\beta}}\nn
  &=&(p,1)^{\otimes 3}\otimes\ketw{\R_{1,(2\cdot\kappa')p'}^{0,\beta}}
\eea
The further analysis of this depends on the parity of $p$ (\ref{ppp}), but having assumed that $p$ is
odd, we use the decomposition of  $\ketw{\Ec,1}$ in (\ref{E1odd}) to obtain
\bea
 &&\Big\{\bigoplus_{\al}^{p-1}(p-\al)\ketw{\R_{2p,1}^{\al,0}}
   \oplus\bigoplus_{\al}^{p-2}(p-\al)
    \ketw{\R_{p,1}^{\al,0}}\Big\}
  \fus\ketw{\R_{1,\kappa' p'}^{0,\beta}}\nn
  &&\qquad\qquad\qquad =\  
   \bigoplus_{\al}^{p-1}(p-\al)\ketw{\R_{2p,\kappa'p'}^{\al,\beta}}
  \oplus\bigoplus_{\al}^{p-2}(p-\al)\ketw{\R_{p,\kappa'p'}^{\al,\beta}}
\eea
We are now faced with yet another disentangling task in order to identify the individual 
fusion products. Since the result of fusing $\big(k\ketw{A}\big)$ with 
$\ketw{\R_{1,\kappa' p'}^{0,\beta}}$ must be divisible by $k$, we find, as $\al$ increases from 
$0$ to $p-1$, that
\bea
 \ketw{\R_{2p,1}^{\al,0}}\fus\ketw{\R_{1,\kappa' p'}^{0,\beta}}
    &=&\ketw{\R_{2p,\kappa'p'}^{\al,\beta}},
   \qquad\qquad\! \al\ \mathrm{even}\nn
 \ketw{\R_{p,1}^{\al,0}}\fus\ketw{\R_{1,\kappa' p'}^{0,\beta}}&=&\ketw{\R_{p,\kappa'p'}^{\al,\beta}},
   \qquad\qquad\ \al\ \mathrm{odd}
\eea
Since $p$ is odd, we also have
\bea
 \ketw{\Oc,1}\fus\ketw{\R_{1,\kappa' p'}^{0,\beta}}
  &=&\lim_{n\to\infty}\Big(\frac{1}{2n}\Big)^3((2n-1)p,1)^{\otimes 3}\otimes
   \ketw{\R_{1,\kappa' p'}^{0,\beta}}\nn
  &=&\lim_{n\to\infty}\Big(\frac{2n-1}{2n}\Big)^3
    (p,1)^{\otimes 3}\otimes\ketw{\R_{1,\kappa'p'}^{0,\beta}}
\eea
that is,
\bea
 &&\Big\{\bigoplus_{\al}^{p-1}(p-\al)
    \ketw{\R_{p,1}^{\al,0}}
  \oplus\bigoplus_{\al}^{p-2}(p-\al)
    \ketw{\R_{2p,1}^{\al,0}}\Big\} \fus\ketw{\R_{1,\kappa' p'}^{0,\beta}}\nn
  &&\qquad\qquad\qquad =\  
   \bigoplus_{\al}^{p-1}(p-\al)\ketw{\R_{p,\kappa'p'}^{\al,\beta}}
  \oplus\bigoplus_{\al}^{p-2}(p-\al)\ketw{\R_{2p,\kappa'p'}^{\al,\beta}}
\eea
from which it follows that
\bea
 \ketw{\R_{p,1}^{\al,0}}\fus\ketw{\R_{1,\kappa' p'}^{0,\beta}}
    &=&\ketw{\R_{p,\kappa'p'}^{\al,\beta}},
   \qquad\qquad\ \al\ \mathrm{even}\nn
 \ketw{\R_{2p,1}^{\al,0}}\fus\ketw{\R_{1,\kappa' p'}^{0,\beta}}
   &=&\ketw{\R_{2p,\kappa'p'}^{\al,\beta}},
   \qquad\qquad \al\ \mathrm{odd}
\eea
Combining these results for general $\al\in\mathbb{Z}_{0,p-1}$, we see that
\be
  \ketw{\R_{\kappa p,1}^{\al,0}}\fus\ketw{\R_{1,\kappa' p'}^{0,\beta}}
    \ =\ \ketw{\R_{\kappa p,\kappa'p'}^{\al,\beta}}
\label{RaRb}
\ee

Returning to the question of self-consistency, it is obvious that one arrives at
the same result (\ref{RaRb}) using (\ref{A1E}) if both $p$ and $p'$ are odd.
For $p$ odd and $p'$ even, self-consistency requires that
\bea
 &&\ketw{\R_{p,1}^{\al,0}}\fus\Big(\bigoplus_{\beta}^{p'-2}(p'-\beta)\Big\{
   \ketw{\R_{1,p'}^{0,\beta}}\oplus\ketw{\R_{1,2p'}^{0,\beta}}\Big\}\Big)\nn
 &=&\ketw{\R_{p,1}^{\al,0}}\fus\ketw{1,\Ec}
  \ =\ \lim_{n\to\infty}\Big(\frac{1}{2n}\Big)^3\ketw{\R_{p,1}^{\al,0}}\otimes(1,2np')^{\otimes 3}\nn
 &=&\ketw{\R_{p,1}^{\al,0}}\otimes
   \Big(\bigoplus_{\beta}^{p'-2}(p'-\beta)
     \Big\{\R_{1,p'}^{0,\beta}\oplus\frac{1}{2}\R_{1,2p'}^{0,\beta}\Big\}\Big)\nn
 &=&\bigoplus_{\beta}^{p'-2}(p'-\beta)
     \Big\{\ketw{\R_{p,p'}^{\al,\beta}}\oplus\ketw{\R_{p,2p'}^{\al,\beta}}\Big\}
\eea
when the left-hand side is evaluated using (\ref{RaRb}). This is easily verified.
Our `symmetric' notation finally ensures that our fusion prescription is self-consistent
also in the case where $p$ is even and $p'$ is odd.

Together with the definitions of ${\cal W}$-indecomposable representations as
simple fusions of ${\cal W}$- and Virasoro-indecomposable representations 
in Section \ref{SectionRepContent}, 
the remarkably simple ${\cal W}$-extended fusion products (\ref{RaRb})
demonstrate that the evaluation of the fusion algebra of ${\cal WLM}(p,p')$ separates into 
horizontal and vertical parts. Furthermore, associativity and commutativity 
of the fusion algebra of ${\cal WLM}(p,p')$ are inherited
from the associative and commutative fundamental fusion algebra of ${\cal LM}(p,p')$ 
in the Virasoro picture.

It is now straightforward to complete the derivation of the fusion algebra of 
${\cal WLM}(p,p')$ as summarized in Section \ref{SectionFusAlg}.
To illustrate this, we first consider
\bea
 \ketw{r,\kappa p'}\fus\ketw{r',\kappa'p'}&=&\Big\{(r,1)\otimes(r',1)\Big\}
  \otimes\Big\{\ketw{1,\kappa p'}\fus\ketw{1,\kappa'p'}\Big\}\nn
  &=&\Big\{\bigoplus_{j=|r-r'|+1,\ \!{\rm by}\ \!2}^{p-|p-r-r'|-1}(j,1)
   \oplus\!\bigoplus_{\al}^{r+r'-p-1}\R_{p,1}^{\al,0}\Big\}\otimes
  \Big\{\bigoplus_{\beta}^{p'-1}\ketw{\R_{1,(\kappa\cdot\kappa')p'}^{0,\beta}}
   \Big\}\nn
 &=&\!\!\bigoplus_{j=|r-r'|+1,\ \!{\rm by}\ \!2}^{p-|p-r-r'|-1}
  \!\!\Big\{\bigoplus_{\beta}^{p'-1}\ketw{\R_{j,(\kappa\cdot\kappa')p'}^{0,\beta}}
    \Big\}
  \oplus\!\bigoplus_{\al}^{r+r'-p-1}\!\!\Big\{
    \bigoplus_{\beta}^{p'-1}\ketw{\R_{\kappa p,\kappa'p'}^{\al,\beta}}
    \Big\}
\eea
in accordance with (\ref{fus11}).
In the second and final example, we consider
\bea
 \ketw{\R_{\kappa p,s}^{a,0}}\fus\ketw{\R_{p,\kappa'p'}^{a',b'}}
  &=&\Big\{\ketw{\R_{\kappa p,1}^{a,0}}\fus\ketw{\R_{p,1}^{a',0}}\Big\}
  \fus\Big\{(1,s)\otimes\ketw{\R_{1,\kappa'p'}^{0,b'}}\Big\}\nn
  &=&\Big\{\bigoplus_{\al}^{p-|a-a'|-1}
     \!\!2\ketw{\R_{\kappa p,1}^{\al,0}}
     \oplus\bigoplus_{\al}^{|p-a-a'|-1}
     \!\!2\ketw{\R_{\kappa p,1}^{\al,0}}\nn
   &&\qquad\oplus\bigoplus_{\al}^{p-|p-a-a'|-1}
     \!\!\!2\ketw{\R_{(2\cdot\kappa)p,1}^{\al,0}}
     \oplus\bigoplus_{\al}^{|a-a'|-1}
     \!2\ketw{\R_{(2\cdot\kappa)p,1}^{\al,0}}\Big\}\nn
  &\fus&\Big\{\bigoplus_{\beta=|b'-s|+1,\ \!{\rm by}\ \!2}^{p'-|p'-s-b'|-1}
      \!\!\!\ketw{\R_{1,\kappa'p'}^{0,\beta}}
   \oplus\bigoplus_{\beta}^{s-b'-1}\!2\ketw{\R_{1,\kappa'p'}^{0,\beta}}
     \oplus\!\!\!\!\bigoplus_{\beta}^{b'+s-p'-1}
        \!\!\!2\ketw{\R_{1,(2\cdot\kappa')p'}^{0,\beta}}\Big\}
    \nn
\eea
By recombining the two components using (\ref{RaRb}), we immediately recognize
the first fusion rule in (\ref{fus23}).

\subsection{Fusion subalgebras without disentanglement}
\label{SectionSub}

There are many fusion subalgebras of the ${\cal W}$-extended fusion algebra (\ref{Wfus}). 
We have already encountered some of them, namely the projective fusion algebra 
discussed in Section
\ref{SectionProj} as well as the horizontal and vertical fusion algebras 
discussed in Section \ref{SectionGeneralp} and Section \ref{SectionVertical}, respectively.
For $p>1$, there is also a six-dimensional fusion subalgebra 
\be
 \big\langle \ketw{\Ec,1},\ketw{\Oc,1},\ketw{1,\Ec},\ketw{1,\Oc},\ketw{\Ac},\ketw{\Bc} \big\rangle_{p,p'}
\label{six}
\ee
where $\ketw{\Ac}$ and $\ketw{\Bc}$ depend on the parities of $p$ and $p'$. To describe
this fusion subalgebra, we introduce the abbreviations
\bea
 \ketw{\Ec,\Ec}:=\ \ketw{\Ec,1}\fus\ketw{1,\Ec},\qquad\quad
   &\ketw{\Ec,\Oc}:=\ \ketw{\Ec,1}\fus\ketw{1,\Oc}\nn
 \ketw{\Oc,\Ec}:=\ \ketw{\Oc,1}\fus\ketw{1,\Ec},\qquad\quad
   &\ketw{\Oc,\Oc}:=\ \ketw{\Oc,1}\fus\ketw{1,\Oc} 
\eea
For $p$ even and $p'$ odd, we then have 
\bea
 \ketw{\Ac}&=&\ketw{\Ec,\Ec}\ =\ \ketw{\Ec,\Oc}
   \ =\ \bigoplus_{\beta=0}^{p'-1}(p'-\beta)\ketw{\Ec,1}\otimes\R_{1,p'}^{0,\beta}\nn
 &=&\bigoplus_{\al}^{p-2}(p-\al)\Big\{\bigoplus_{\beta=0}^{p'-1}(p'-\beta)
   \Big( \ketw{\R_{p,p'}^{\al,\beta}}\oplus\ketw{\R_{2p,p'}^{\al,\beta}}\Big)\Big\}    \nn
 \ketw{\Bc}&=&\ketw{\Oc,\Ec}\ =\ \ketw{\Oc,\Oc}
   \ =\ \bigoplus_{\beta=0}^{p'-1}(p'-\beta)\ketw{\Oc,1}\otimes\R_{1,p'}^{0,\beta}\nn
 &=&\bigoplus_{\al}^{p-1}(p-\al)\Big\{\bigoplus_{\beta=0}^{p'-1}(p'-\beta)
   \Big( \ketw{\R_{p,p'}^{\al,\beta}}\oplus\ketw{\R_{2p,p'}^{\al,\beta}}\Big)\Big\}  
\eea
For $p$ odd and $p'$ even, we have 
\bea
 \ketw{\Ac}&=&\ketw{\Ec,\Ec}\ =\ \ketw{\Oc,\Ec}
    \ =\ \bigoplus_{\beta}^{p'-2}
     (p'-\beta)\Big\{\ketw{\Ec,1}\oplus\ketw{\Oc,1}\Big\}\otimes\R_{1,p'}^{0,\beta}\nn
   &=& \bigoplus_{\beta}^{p'-2}(p'-\beta)\Big\{\bigoplus_{\al=0}^{p-1}(p-\al)
   \Big( \ketw{\R_{p,p'}^{\al,\beta}}\oplus\ketw{\R_{2p,p'}^{\al,\beta}}\Big)\Big\}   \nn
 \ketw{\Bc}&=&\ketw{\Ec,\Oc}\ =\ \ketw{\Oc,\Oc}
   \ =\ \bigoplus_{\beta}^{p'-1}(p'-\beta)
   \Big\{\ketw{\Ec,1}\oplus\ketw{\Oc,1}\Big\}\otimes\R_{1,p'}^{0,\beta}\nn
  &=& \bigoplus_{\beta}^{p'-1}(p'-\beta)\Big\{\bigoplus_{\al=0}^{p-1}(p-\al)
   \Big( \ketw{\R_{p,p'}^{\al,\beta}}\oplus\ketw{\R_{2p,p'}^{\al,\beta}}\Big)\Big\} 
\eea
Finally, for $p$ and $p'$ both odd, we have 
\bea
 \ketw{\Ac}&=&\ketw{\Ec,\Ec}\ =\ \ketw{\Oc,\Oc}
  \ =\ \Big\{\bigoplus_{\beta}^{p'-2}
     (p'-\beta)\ketw{\Ec,1}\otimes\R_{1,p'}^{0,\beta}\Big\}\oplus
  \Big\{\bigoplus_{\beta}^{p'-1}(p'-\beta)\ketw{\Oc,1}\otimes\R_{1,p'}^{0,\beta}\Big\}\nn
 &=& \bigoplus_{\al}^{p-1}(p-\al)\Big\{\bigoplus_{\beta}^{p'-1}(p'-\beta)\ketw{\R_{p,p'}^{\al,\beta}}\Big\}
  \oplus
   \bigoplus_{\al}^{p-2}(p-\al)\Big\{\bigoplus_{\beta}^{p'-2}(p'-\beta)\ketw{\R_{p,p'}^{\al,\beta}}\Big\}\nn
 &&\qquad\oplus 
   \bigoplus_{\al}^{p-1}(p-\al)\Big\{\bigoplus_{\beta}^{p'-2}(p'-\beta)\ketw{\R_{2p,p'}^{\al,\beta}}\Big\}
  \oplus
   \bigoplus_{\al}^{p-2}(p-\al)\Big\{\bigoplus_{\beta}^{p'-1}(p'-\beta)\ketw{\R_{2p,p'}^{\al,\beta}}\Big\}\nn
 \ketw{\Bc}&=&\ketw{\Ec,\Oc}\ =\ \ketw{\Oc,\Ec}
  \ =\ \Big\{\bigoplus_{\beta}^{p'-1}
     (p'-\beta)\ketw{\Ec,1}\otimes\R_{1,p'}^{0,\beta}\Big\}\oplus
  \Big\{\bigoplus_{\beta}^{p'-2}(p'-\beta)\ketw{\Oc,1}\otimes\R_{1,p'}^{0,\beta}\Big\}\nn
 &=& \bigoplus_{\al}^{p-1}(p-\al)\Big\{\bigoplus_{\beta}^{p'-2}(p'-\beta)\ketw{\R_{p,p'}^{\al,\beta}}\Big\}
  \oplus
   \bigoplus_{\al}^{p-2}(p-\al)\Big\{\bigoplus_{\beta}^{p'-1}(p'-\beta)\ketw{\R_{p,p'}^{\al,\beta}}\Big\}\nn
 &&\qquad\oplus 
   \bigoplus_{\al}^{p-1}(p-\al)\Big\{\bigoplus_{\beta}^{p'-1}(p'-\beta)\ketw{\R_{2p,p'}^{\al,\beta}}\Big\}
  \oplus
   \bigoplus_{\al}^{p-2}(p-\al)\Big\{\bigoplus_{\beta}^{p'-2}(p'-\beta)\ketw{\R_{2p,p'}^{\al,\beta}}\Big\}
\eea
\psset{unit=1cm}
\begin{figure}
$$
\renewcommand{\arraystretch}{1.5}
\begin{array}{c||cc|cc|cc}
\hat\otimes&\ketw{\Ec,1}&\ketw{\Oc,1}&\ketw{1,\Ec}&\ketw{1,\Oc}&\ketw{\Ec,\Ec}&\ketw{\Oc,\Oc}\\[4pt]
\hline \hline
\rule{0pt}{14pt}
 \ketw{\Ec,1}&p^3\ketw{\Oc,1}&p^3\ketw{\Ec,1}&\ketw{\Ec,\Ec}&\ketw{\Ec,\Ec}
    &p^3\ketw{\Oc,\Oc}&p^3\ketw{\Ec,\Ec}\\[4pt]
 \ketw{\Oc,1}&p^3\ketw{\Ec,1}&p^3\ketw{\Oc,1}&\ketw{\Oc,\Oc}&\ketw{\Oc,\Oc}
    &p^3\ketw{\Ec,\Ec}&p^3\ketw{\Oc,\Oc}\\[4pt]
\hline
\rule{0pt}{14pt}
 \ketw{1,\Ec}&\ketw{\Ec,\Ec}&\ketw{\Oc,\Oc}&p'^3\ketw{1,\Oc}&p'^3\ketw{1,\Ec}
   &p'^3\ketw{\Ec,\Ec}&p'^3\ketw{\Oc,\Oc}\\[4pt]
 \ketw{1,\Oc}&\ketw{\Ec,\Ec}&\ketw{\Oc,\Oc}&p'^3\ketw{1,\Ec}&p'^3\ketw{1,\Oc}
   &p'^3\ketw{\Ec,\Ec}&p'^3\ketw{\Oc,\Oc}\\[4pt]
\hline
\rule{0pt}{14pt}
 \ketw{\Ec,\Ec}&p^3\ketw{\Oc,\Oc}&p^3\ketw{\Ec,\Ec}&p'^3\ketw{\Ec,\Ec}&p'^3\ketw{\Ec,\Ec}
   &(pp')^3\ketw{\Oc,\Oc}&(pp')^3\ketw{\Ec,\Ec}\\[4pt]
 \ketw{\Oc,\Oc}&p^3\ketw{\Ec,\Ec}&p^3\ketw{\Oc,\Oc}&p'^3\ketw{\Oc,\Oc}&p'^3\ketw{\Oc,\Oc}
   &(pp')^3\ketw{\Ec,\Ec}&(pp')^3\ketw{\Oc,\Oc}
\end{array}
$$
\caption{Cayley table of the six-dimensional $\Ec,\Oc$ fusion subalgebra (\ref{six})
for $p$ even and $p'$ odd.}
\label{EOevenodd}
\end{figure}
\psset{unit=1cm}
\begin{figure}
$$
\renewcommand{\arraystretch}{1.5}
\begin{array}{c||cc|cc|cc}
\hat\otimes&\ketw{\Ec,1}&\ketw{\Oc,1}&\ketw{1,\Ec}&\ketw{1,\Oc}&\ketw{\Ec,\Ec}&\ketw{\Oc,\Oc}\\[4pt]
\hline \hline
\rule{0pt}{14pt}
 \ketw{\Ec,1}&p^3\ketw{\Oc,1}&p^3\ketw{\Ec,1}&\ketw{\Ec,\Ec}&\ketw{\Oc,\Oc}
    &p^3\ketw{\Ec,\Ec}&p^3\ketw{\Oc,\Oc}\\[4pt]
 \ketw{\Oc,1}&p^3\ketw{\Ec,1}&p^3\ketw{\Oc,1}&\ketw{\Ec,\Ec}&\ketw{\Oc,\Oc}
    &p^3\ketw{\Ec,\Ec}&p^3\ketw{\Oc,\Oc}\\[4pt]
\hline
\rule{0pt}{14pt}
 \ketw{1,\Ec}&\ketw{\Ec,\Ec}&\ketw{\Ec,\Ec}&p'^3\ketw{1,\Oc}&p'^3\ketw{1,\Ec}
    &p'^3\ketw{\Oc,\Oc}&p'^3\ketw{\Ec,\Ec}\\[4pt]
 \ketw{1,\Oc}&\ketw{\Oc,\Oc}&\ketw{\Oc,\Oc}&p'^3\ketw{1,\Ec}&p'^3\ketw{1,\Oc}
    &p'^3\ketw{\Ec,\Ec}&p'^3\ketw{\Oc,\Oc}\\[4pt]
\hline
\rule{0pt}{14pt}
 \ketw{\Ec,\Ec}&p^3\ketw{\Ec,\Ec}&p^3\ketw{\Ec,\Ec}&p'^3\ketw{\Oc,\Oc}&p'^3\ketw{\Ec,\Ec}
    &(pp')^3\ketw{\Oc,\Oc}&(pp')^3\ketw{\Ec,\Ec}\\[4pt]
 \ketw{\Oc,\Oc}&p^3\ketw{\Oc,\Oc}&p^3\ketw{\Oc,\Oc}&p'^3\ketw{\Ec,\Ec}
    &p'^3\ketw{\Oc,\Oc}&(pp')^3\ketw{\Ec,\Ec}&(pp')^3\ketw{\Oc,\Oc}
\end{array}
$$
\caption{Cayley table of the six-dimensional $\Ec,\Oc$ fusion subalgebra (\ref{six})
for $p$ odd and $p'$ even.}
\label{EOoddeven}
\end{figure}
\psset{unit=1cm}
\begin{figure}
$$
\renewcommand{\arraystretch}{1.5}
\begin{array}{c||cc|cc|cc}
\hat\otimes&\ketw{\Ec,1}&\ketw{\Oc,1}&\ketw{1,\Ec}&\ketw{1,\Oc}&\ketw{\Ec,\Ec}&\ketw{\Oc,\Ec}\\[4pt]
\hline \hline
\rule{0pt}{14pt}
 \ketw{\Ec,1}&p^3\ketw{\Oc,1}&p^3\ketw{\Ec,1}&\ketw{\Ec,\Ec}&\ketw{\Oc,\Ec}
    &p^3\ketw{\Oc,\Ec}&p^3\ketw{\Ec,\Ec}\\[4pt]
 \ketw{\Oc,1}&p^3\ketw{\Ec,1}&p^3\ketw{\Oc,1}&\ketw{\Oc,\Ec}&\ketw{\Ec,\Ec}
    &p^3\ketw{\Ec,\Ec}&p^3\ketw{\Oc,\Ec}\\[4pt]
\hline
\rule{0pt}{14pt}
 \ketw{1,\Ec}&\ketw{\Ec,\Ec}&\ketw{\Oc,\Ec}&p'^3\ketw{1,\Oc}&p'^3\ketw{1,\Ec}
    &p'^3\ketw{\Oc,\Ec}&p'^3\ketw{\Ec,\Ec}\\[4pt]
 \ketw{1,\Oc}&\ketw{\Oc,\Ec}&\ketw{\Ec,\Ec}&p'^3\ketw{1,\Ec}&p'^3\ketw{1,\Oc}
    &p'^3\ketw{\Ec,\Ec}&p'^3\ketw{\Oc,\Ec}\\[4pt]
\hline
\rule{0pt}{14pt}
 \ketw{\Ec,\Ec}&p^3\ketw{\Oc,\Ec}&p^3\ketw{\Ec,\Ec}&p'^3\ketw{\Oc,\Ec}
    &p'^3\ketw{\Ec,\Ec}&(pp')^3\ketw{\Ec,\Ec}&(pp')^3\ketw{\Oc,\Ec}\\[4pt]
 \ketw{\Oc,\Ec}&p^3\ketw{\Ec,\Ec}&p^3\ketw{\Oc,\Ec}&p'^3\ketw{\Ec,\Ec}
    &p'^3\ketw{\Oc,\Ec}&(pp')^3\ketw{\Oc,\Ec}&(pp')^3\ketw{\Ec,\Ec}
\end{array}
$$
\caption{Cayley table of the six-dimensional $\Ec,\Oc$ fusion subalgebra (\ref{six})
for $p$ and $p'$ both odd.}
\label{EOoddodd}
\end{figure}
We note that the linear relations discussed in Section \ref{SectionLinRel} simply correspond
to the identities arising when alternatively decomposing the ${\cal W}$-representations
$\ketw{\Ac}$ and $\ketw{\Bc}$ in terms of fusions involving the {\em vertical} ${\cal W}$-representations
$\ketw{1,\Ec}$ and $\ketw{1,\Oc}$.
For the various parities of $p$ and $p'$, these six-dimensional fusion algebras are indicated in 
Figure \ref{EOevenodd}, Figure \ref{EOoddeven} and Figure \ref{EOoddodd}.
Besides the two-dimensional fusion subalgebras, the horizontal subalgebra
$\big\langle\ketw{\Ec,1},\ketw{\Oc,1}\big\rangle_{p,p'}$ and the vertical subalgebra
$\big\langle\ketw{1,\Ec},\ketw{1,\Oc}\big\rangle_{p,p'}$, we note the additional two-dimensional
fusion subalgebra $\big\langle\ketw{\Ac},\ketw{\Bc}\big\rangle_{p,p'}$.
All three of these two-dimensional fusion subalgebras are easily identified
along the diagonals of the Cayley tables in the three figures. 

We excluded $p=1$ in the discussion of the six-dimensional fusion subalgebra.
This was necessitated by the fact that for $p=1$, the six ${\cal W}$-representations 
in (\ref{six}) are linearly dependent
\bea
 &&\ketw{\Ac}\ =\ \ketw{1,\Ec},\qquad
 \ketw{\Bc}\ =\ \ketw{1,\Oc},\qquad\quad p'\ \mathrm{even}\nn
 &&\ketw{\Ac}\ =\ \ketw{1,\Oc},\qquad
 \ketw{\Bc}\ =\ \ketw{1,\Ec},\qquad\quad p'\ \mathrm{odd}
\eea
The decompositions of these vertical representations given in 
(\ref{1Eeven}) for $p'$ even and in (\ref{1Eodd}) for $p'$ odd are unaffected by setting $p=1$.
Thus, there is a four-dimensional fusion subalgebra 
\be
 \big\langle \ketw{1,1},\ketw{2,1},\ketw{1,\Ec},\ketw{1,\Oc}\big\rangle_{1,p'}
\label{four}
\ee
of the ${\cal W}$-extended fusion algebra of ${\cal WLM}(1,p')$, where
\be
 \ketw{\Ec,1}\ =\ \ketw{2,1},\qquad\quad\ketw{\Oc,1}\ =\ \ketw{1,1}
\ee
The fusion rules governing (\ref{four}) are given in the Cayley table in Figure \ref{Table1peven} for $p'$ 
even and in the Cayley table in Figure \ref{Table1podd} for $p'$ odd. 

We finally stress that, for every ${\cal W}$-extended logarithmic minimal model ${\cal WLM}(p,p')$,
a virtue of the six-dimensional (or four-dimensional for $p=1$) 
fusion subalgebra just described is that it does {\em not} rely on any disentangling procedure.
\psset{unit=1cm}
\begin{figure}
$$
\renewcommand{\arraystretch}{1.5}
\begin{array}{c||cc|cc}
\hat\otimes&\ketw{1,1}&\ketw{2,1}&\ketw{1,\Oc}&\ketw{1,\Ec}\\[4pt]
\hline \hline
\rule{0pt}{14pt}
 \ketw{1,1}&\ketw{1,1}&\ketw{2,1}&\ketw{1,\Oc}&\ketw{1,\Ec}
    \\[4pt]
 \ketw{2,1}&\ketw{2,1}&\ketw{1,1}&\ketw{1,\Oc}&\ketw{1,\Ec}
    \\[4pt]
\hline
\rule{0pt}{14pt}
 \ketw{1,\Oc}&\ketw{1,\Oc}&\ketw{1,\Oc}&p'^3\ketw{1,\Oc}&p'^3\ketw{1,\Ec}
    \\[4pt]
 \ketw{1,\Ec}&\ketw{1,\Ec}&\ketw{1,\Ec}&p'^3\ketw{1,\Ec}&p'^3\ketw{1,\Oc}
\end{array}
$$
\caption{Cayley table of the four-dimensional $\Ec,\Oc$ fusion subalgebra (\ref{four}) for $p'$ even.}
\label{Table1peven}
\end{figure}
\psset{unit=1cm}
\begin{figure}
$$
\renewcommand{\arraystretch}{1.5}
\begin{array}{c||cc|cc}
\hat\otimes&\ketw{1,1}&\ketw{2,1}&\ketw{1,\Oc}&\ketw{1,\Ec}\\[4pt]
\hline \hline
\rule{0pt}{14pt}
 \ketw{1,1}&\ketw{1,1}&\ketw{2,1}&\ketw{1,\Oc}&\ketw{1,\Ec}
    \\[4pt]
 \ketw{2,1}&\ketw{2,1}&\ketw{1,1}&\ketw{1,\Ec}&\ketw{1,\Oc}
    \\[4pt]
\hline
\rule{0pt}{14pt}
 \ketw{1,\Oc}&\ketw{1,\Oc}&\ketw{1,\Ec}&p'^3\ketw{1,\Oc}&p'^3\ketw{1,\Ec}
    \\[4pt]
 \ketw{1,\Ec}&\ketw{1,\Ec}&\ketw{1,\Oc}&p'^3\ketw{1,\Ec}&p'^3\ketw{1,\Oc}
\end{array}
$$
\caption{Cayley table of the four-dimensional $\Ec,\Oc$ fusion subalgebra (\ref{four}) for $p'$ odd.}
\label{Table1podd}
\end{figure}

\section{Discussion}
\label{SectionDiscussion}

There is an infinite series of Yang-Baxter integrable logarithmic minimal 
models ${\cal LM}(p,p')$~\cite{PRZ0607}. 
As in the rational case~\cite{BP01}, the Yang-Baxter integrable boundary 
conditions give insight into the conformal boundary conditions~\cite{BPPZ00} 
in the continuum scaling limit as well as into the fusion of their associated Virasoro representations.
This enabled us in~\cite{PRZ0607} to construct
integrable boundary conditions labelled by $(r,s)$ and corresponding to 
so-called Kac representations with
conformal weights in an infinitely extended Kac table.
Moreover, from the lattice implementation of fusion, we obtained~\cite{RP0706,RP0707} 
the closed (fundamental) fusion algebra generated
by these Kac representations finding that indecomposable representations of ranks 1, 2 and 3 are
generated by the fusion process. In the special case where $p=1$, only 
indecomposable representations of rank 1 or 2 arise.
Although there is a countable infinity of representations for general ${\cal LM}(p,p')$,
the ensuing fusion rules are quasi-rational in the sense of Nahm~\cite{Nahm94}, that is,
the fusion of any two indecomposable representations decomposes into a finite sum of 
indecomposable representations.
This is the relevant picture in the case where the conformal algebra is the Virasoro algebra.
Of course, there is no claim, in the context of this logarithmic CFT, that the representations
generated in this picture exhaust all of the representations associated with conformal 
boundary conditions.
This is in stark contrast to the situation in rational CFTs where all representations 
decompose into direct sums of a finite number of irreducible representations.

In this paper, we have reconsidered the lattice description of the logarithmic minimal model
${\cal LM}(p,p')$ in the continuum scaling limit to expose its nature as a `rational' logarithmic CFT
with respect to a ${\cal W}$-extended conformal algebra. Under the extended symmetry, the infinity of
Virasoro representations are reorganized into a finite number of ${\cal W}$-representations.
Following the approach of \cite{PRR0803,RP0804}, we have constructed new solutions of the boundary 
Yang-Baxter equation which, in a particular limit, correspond to these representations. 
With respect to a suitably defined ${\cal W}$-fusion implemented on the lattice, we find that the
representation content of the ensuing closed, associative and commutative fusion algebra is {\em finite}
containing $6pp'-2p-2p'$ ${\cal W}$-indecomposable representations with 
$2p+2p'-2$ rank-1 representations,
$4pp'-2p-2p'$ rank-2 representations and $2(p-1)(p'-1)$ rank-3 representations. 
The ${\cal W}$-indecomposable rank-1 representations are all ${\cal W}$-irreducible while 
we have presented a conjecture for the embedding patterns of the ${\cal W}$-indecomposable
rank-2 and -3 representations.
We have also identified their
associated ${\cal W}$-extended characters which decompose as finite non-negative sums
of $2pp'+(p-1)(p'-1)/2$ distinct ${\cal W}$-irreducible characters. 
For $2p+2p'-4$ of the rank-2 and all of the rank-3 ${\cal W}$-indecomposable representations,
we have presented fermionic character expressions.
To distinguish between
inequivalent ${\cal W}$-indecomposable representations of identical characters,
we have introduced `refined' characters carrying information also about the Jordan-cell content 
of a representation. Furthermore, we have found that 
$2pp'$ of the ${\cal W}$-indecomposable representations are in fact
${\cal W}$-projective representations and shown that they generate a closed fusion subalgebra.
Finally, we interpret the closure of the ${\cal W}$-indecomposable representations among 
themselves under fusion as confirmation of the proposed extended symmetry.

The results presented in this paper apply to the entire infinite series ${\cal WLM}(p,p')$.
Some of these models are of great interest and have been studied before. In particular,
symplectic fermions ${\cal WLM}(1,2)$ (which are critical dense polymers ${\cal LM}(1,2)$
viewed in the ${\cal W}$-extended picture~\cite{PRR0803}) and more generally
the infinite series ${\cal WLM}(1,p')$ are discussed in 
\cite{GK9606,FHST03,FGST05,GR07,GTipunin07,PRR0803}, while 
${\cal W}$-extended critical percolation ${\cal WLM}(2,3)$ is discussed in \cite{RP0804}.
One may verify explicitly that our general expressions for characters and fusion rules
indeed reduce to the expressions given in those papers when
fixing $p$ and $p'$ to their relevant values. Among the many other interesting models
are the ${\cal W}$-extended logarithmic Yang-Lee model ${\cal WLM}(2,5)$ and
the ${\cal W}$-extended logarithmic Ising model ${\cal WLM}(3,4)$.
The numbers of ${\cal W}$-indecomposable and ${\cal W}$-irreducible representations
are rather large as they are given by
\bea
 N_1(2,5)\ =\ 12,\quad N_2(2,5)\ =\ 26,\quad N_3(2,5)\ =\ 8,\ \ \!\quad N_{\rm{ind}}(2,5)\ =\ 46,\quad
  N_{\mathrm{irr}}(2,5)\ =\ 22\nn
 N_1(3,4)\ =\ 12,\quad N_2(3,4)\ =\ 34,\quad N_3(3,4)\ =\ 12,\quad N_{\rm{ind}}(3,4)\ =\ 58,\quad
  N_{\mathrm{irr}}(3,4)\ =\ 27
\eea 
while the numbers of ${\cal W}$-projective representations are
\be
 N_{\mathrm{proj}}(2,5)\ =\ 20,\qquad N_{\mathrm{proj}}(3,4)\ =\ 24
\ee

A somewhat surprising feature of our closed ${\cal W}$-extended fusion algebra of 
${\cal WLM}(p,p')$ is that there appears
to be no natural identity $\Ic_{\cal W}$ expressed in terms of the fundamental Virasoro fusion
algebra and with respect to the fusion multiplication $\hat\otimes$.
Since the Kac representation $(1,1)$ is the identity
of the fundamental fusion algebra itself, it may be tempting to include it in the ${\cal W}$-extended
spectrum and identify it with $\Ic_{\cal W}$. However, we have
\be
 \ketw{\Ec,1}\fus\Ic_{\cal W}:=\ 
  \lim_{n\to\infty}\Big(\frac{1}{2n}\Big)^3(2np,1)^{\otimes3}\otimes(1,1)\ =\ 0 
\ee
demonstrating that this simple extension fails.
We find it natural, though, to expect that one can extend our fusion algebra
of ${\cal WLM}(p,p')$ by working with the {\em full} Virasoro fusion algebra.
We hope to discuss this and re-address the identity question elsewhere.

Comparing the sets
of  ${\cal W}$-irreducible and ${\cal W}$-indecomposable rank-1 representations
with the results of~\cite{FGST06b}, we find complete agreement.
In a further comparison, the sets of ${\cal W}$-projective characters agree as well.
We therefore find it natural to suspect that our construction is with respect
to their extended conformal algebra ${\cal W}_{p,p'}$. 
Here we wish to point out that these ${\cal W}$-projective characters were found
to constitute a representation of the modular group in \cite{FGST06b}
and that several modular invariants can be formed out of these.
Combining this with our observation that the corresponding ${\cal W}$-projective {\em representations}
generate a closed fusion algebra, yields an intriguing hint towards the classification
of torus amplitudes in ${\cal WLM}(p,p')$.
We also wish to emphasize that the works \cite{FGST06b,Sem0710} address fusion {\em only} 
by studying the Grothendieck ring of a related quantum group
as an {\em approximation} to the fusion algebra
of their $2pp'$ representations ${\cal K}_{r,s}^{\pm}$. In this context, the Grothendieck
ring may be regarded as the `fusion algebra' of the corresponding set of ${\cal W}$-{\em characters} 
as opposed to the much richer fusion algebra of ${\cal W}$-{\em representations}.
The latter is given explicitly in Section \ref{SectionFusAlg} above and is one of our main results.
%
%
\section*{Acknowledgments}
\vskip.1cm
\noindent
This work is supported by the Australian Research Council (ARC). 
The author thanks Paul A. Pearce and Ilya Yu. \!Tipunin for useful discussions and comments.

\end{document}